\documentclass{jpsj2}

\def\runauthor{Online-Journal Subcommittee}

\title{%
Fate of Quasiparticle at Mott Transition and Interplay with Lifshitz Transition Studied by Correlator Projection Method}

\author{%
Kota {\sc Hanasaki}$^{1}$ and Masatoshi {\sc Imada}$^{1,2,3}$
}

\inst{%
$^{1}$Institute for Solid State Physics, University of Tokyo, Kashiwanoha 5-1-5, Kashiwa, Chiba 277-8581\\
$^{2}$Department of Applied Physics, University of Tokyo, Hongo 7-3-1, Bunkyo-ku, Tokyo\\
$^{3}$PRESTO, Japan Science and Technology Agency
}

\abst{%
Filling-control metal-insulator transition on the two-dimensional Hubbard model is investigated by using the correlator projection method, which takes into account momentum dependence of the free energy beyond the dynamical mean-field theory. 
The phase diagram of metals and Mott insulators is analyzed. Lifshitz transitions occur simultaneously with metal-insulator transitions at large Coulomb repulsion. On the other hand, they are separated each other for lower Coulomb repulsion, where the phase sandwiched by the Lifshitz and metal-insulator transitions appears to show violation of the Luttinger sum rule. 
Through the metal-insulator transition, quasiparticles retain nonzero renormalization factor and finite quasi-particle weight in the both sides of the transition.  This supports that the metal-insulator transition is caused not by the vanishing renormalization factor but by the relative shift of the Fermi level into the Mott gap away from the {\it quasiparticle band}, in sharp contrast with the original dynamical mean-field theory.
Charge compressibility diverges at the critical end point of the first-order Lifshitz transition at finite temperatures. The origin of the divergence is ascribed to singular momentum dependence of the quasiparticle dispersion.
}

\kword{%
Filling-control, metal-insulator transition, Hubbard model, Lifshitz transition,compressibility divergence, quasiparticle dispersion
}

\begin{document}
\maketitle
\section{Introduction}

Transitions between metals and insulators caused by electron correlation effects have been studied extensively\cite{IFT-MIT}.  A typical example of the correlation-driven metal-insulator transitions is the Mott transition, near which various competing orders including unconventional superconductivity in the copper oxides are found.  In spite of long history, the nature of Mott transitions still remains a challenge of recent active research and a controversy remains.  
A key issue of the Mott transition\cite{IFT-MIT} has already been posed by early studies by Hubbard\cite{Hubbard1} and Brinkman-Rice\cite{BrinkmanRice}, which have proposed somewhat contradicting scenarios each other. The former describes the formation of energy gap and the insulator appears from the shift of the Fermi level into the gap, while the latter describes the transition to the insulator by the mass divergence of the quasiparticle. 

Recently, the two views are unified in the limit of infinite dimension by the dynamical mean-field theory (DMFT)\cite{GKKR-DMFT,MITDMFT}. DMFT succeeded in describing the two aspects of the Mott transition; formation of an energy gap and vanishing of the Kondo-like resonance at the Fermi level. The latter is described as vanishing quasiparticle weight at the Fermi level $Z \to 0$. Although the Mott gap is formed already in the metallic phase before the transition, the criticality of the metal-insulator transition is governed by the vanishing renormalization factor of the coherent band at the Fermi level.

Unfortunately, however, DMFT is a theory which is exact only in the infinite-dimension limit and its validity in finite dimensions is not clear. In finite dimensions, the self-energy becomes momentum dependent in contrast to DMFT. 
Results of recent angle-resolved photoemission spectroscopy (ARPES) measurements\cite{Yoshida1,Yoshida2} have revealed momentum dependent aspects of metal-insulator transition (MIT) in low-dimensional systems like cupurate superconductors. In these systems, momentum space anisotropy of quasiparticles is especially enhanced in low-doping samples, which implies that the momentum dependence of the self-energy may play a crucial role in MIT and physics around it.
Therefore, it is desired to extend DMFT to allow the momentum dependence and clarify how it modifies the nature of the transition.

The central issue is whether the Mott transition in finite dimensions is caused by the vanishing of the quasiparticle at the Fermi level or caused by the move of the quasiparticle dispersion out of the Fermi level. The answer to this fundamental question may depend on the way of treating the momentum dependence of quasiparticles and therefore one has to be careful in obtaining reliable answer in finite dimensions.  In this paper, we study this issue by considering the momentum dependence of the self-energy beyond the dynamical mean-field theory and clarify the fate of quasiparticle in realistic finite dimensional systems.  

We reiterate the two possible scenarios of the Mott transition in more detail in the following:

\noindent{(1) {\it Quasiparticle rigid-band picture}}

\noindent{It is based on a certain `band structure' of quasiparticles.}
Here, the `band structure' does not necessarily mean the non-interacting one. It is the energy dispersion of quasiparticles, which may have gap-like structure formed by electron-electron correlations.
MIT is described as Fermi level shift out of the top/bottom of the quasiparticle band. Namely, Fermi level excitations disappear because the Fermi level moves to the energy range of the gap where there are no quasiparticles. 
Its prototypical example is the Hubbard picture of MIT\cite{Hubbard1}, although the Hubbard approximation itself clearly oversimplifies the real transition.
If this route is realized, at least the quasiparticle weight closest to the Fermi level remains nonzero all through the transition.

\noindent{(2) {\it Quasiparticle-weight vanishing picture} }

\noindent{The other type of description is similar to the Brinkman-Rice scenario~\cite{BrinkmanRice} stating that the weight of the quasiparticle disappears at the Fermi level.}
Quasiparticle poles stay at the Fermi level by keeping the Fermi surface until the transition but their weights disappear at the transition.
The MIT is purely driven by the weight disappearance. We categorize the DMFT scenario of MIT here since it describes the MIT as a formation and a destruction of a resonant excitation at the Fermi level~\cite{FCDMFT}.

We note that the real transition may occur in a compromized manner and may have the both characters. In other words, in reallity, both of the gap formation and the weight renormalization may occur simultaneously. However, there still remains a crucial issue whether the quasiparticle weight remains or not at MIT. In its criticality, the two routes are not compatible at least when the transition is continuous without the phase separation.

Another related issue in two-dimensional systems is the relationship of the Mott transition to the Lifshitz transition.  If the Mott gap is preformed in the metallic phase, the Lifshitz transition may trigger a transition from large Fermi surface to small pockets through the arc-like structure. However, the first-order Lifshitz transition may preempt this phase with the small pockets and cause a simultaneous Lifshitz and metal-insulator transition.  The fate of the quasiparticle might depend on these different choices of the routes.  Such an issue is relevant only in finite dimensional systems and has not been explored in detail.

Recently there have been several attempts of momentum-sensitive analysis beyond DMFT. The cellular-DMFT (cDMFT)~\cite{cDMFT}, the dynamical cluster approximation (DCA)~\cite{DCA} and the cluster pertubation theory (CPT)~\cite{CPT} have been applied with certain successes. 
However, the resolution of analyzing spatial fluctuations is limited because it is difficult to increase the number of clusters in space. This makes it difficult to discuss growth of short-ranged correlations. 
Here instead, we use the correlator projection method (CPM)~\cite{OnodaPRB}, where spatial fluctuations can be considered with a large number of momentum points.  From the sufficient number of momenta in the Brillouin zone, one can discuss, for example, evolution of the Fermi surface near the metal-insulator transition point and momentum dependence of quasiparticle properties in detail.  Our scheme interpolates high and low energy regions by taking equation of motion method combined with the improved DMFT.  The high-energy part is systematically taken into account by the continued fraction derived from the equation of motion, while the low-energy physics is incorporated in the self-consistent scheme of DMFT but here for the self-consistent determination of the higher-order self-energy, which now has the momentum dependence in contrast to the truncation at the lowest-order self-energy in the original DMFT.   

In order to investigate the fundamental issues we have mentioned, we employ the two dimensional Hubbard model with next-nearest-neighbor hopping as is defined by
\begin{equation}
{\cal H}= -\sum_{\sigma}\sum_{(i,j):{\rm n.n.}}t c^{\dagger}_{i\sigma}c_{j\sigma}-\sum_{\sigma}\sum_{(i,j):{\rm n.n.n.}}t^{\prime} c^{\dagger}_{i\sigma}c_{j\sigma}
+ U\sum_{i}n_{i\uparrow}n_{i\downarrow}.
\label{eqn:HM}
\end{equation}
 Here, $c_{i\sigma}(c^{\dagger}_{i\sigma})$ is the annihilation (creation) operator of an electron at an atomic site $i$ with a spin index $\sigma$. The number operator on site $i$, spin $\sigma$ is denoted by $n_{i,\sigma}$.
The local Coulomb repulsion is denoted by $U$.
The parameter $t$ and $t^{\prime}$ are the electron transfers between the nearest neighbor (n.n.) sites and those between the next nearest neighbor (n.n.n.) sites, respectively. 
The summations $\sum_{(i,j):{\rm n.n.}}$ and $\sum_{(i,j):{\rm n.n.n.}}$ are taken for all the pairs $(i,j)$ between the nearest neighbor or between the next nearest neighbor sites, respectively. 
Here, we set n.n.n. hopping parameter $t^{\prime}$ as $t^{\prime}=-0.20t$.
The properties of this model at this value of $t^{\prime}$ have been investigated with several methods.
At zero temperature, bandwidth-control (BC) MIT as well as filling-control (FC) MIT has been investigated using the path integral renormalization group method (PIRG),~\cite{KashimaPIRG,WatanabePIRG} where the first-order metal-insulator transition occurs at a finite value of $U$.
At finite temperature, BC MIT has been studied with the correlator projection method (CPM)~\cite{OnodaPRB}, which indicates that the first-order MIT extends to finite temperatures and have a critical point at the end of the first-order transition.
Finite-temperature study of FC MIT is not available and is of great interest.   

In Sec.2, we briefly summarize the formulation of the correlator projection method. Numerical results are presented in Sec.3. Section 4 is devoted to discussions and conclusion.

\section{Formulation}
\subsection{Standard Formulation for bandwidth-control problems}
The standard formulation of CPM (for BC problems) is given in previous papers\cite{OnodaPRB, OnodaJPSJ}. Here, we just summarize the basic formalism.

First the equation of motions for operators are solved to give a continued fraction expansion of the Green's function. We start from a set of operators $\{ c_{i\sigma} \}$ as the first operator set, and introduce the second and the third operator set as $\{ \phi^{(2)}_{i\sigma} \}$ and  $\{ \phi^{(3)}_{i\sigma} \}$, respectively. These operators are defined through successive operator projection procedure. 
Here the projection of an arbitrary operator $O$ to the n-th operator set $\{\phi^{(n)}\}$ is defined as
\[  \hat{P}_{n}O \equiv \sum_{l}\langle \{ O, \phi^{(n)}_{l}\}\rangle \left( S^{(n) -1} \right)_{lm}\phi^{(n)}_{m}, \]
where $\{\cdot,\cdot\}$ is an anticommutator and $\langle \cdot \cdot \cdot \rangle$ is the thermodynamic average. The overlap matrix for the n-th operator set, $\langle \{ \phi^{(n)}_{l}, \phi^{(n)}_{m}\} \rangle$, is denoted by $S^{(n)}_{lm}$.
Then the higher-order operators are defined as
\[ \phi^{(n+1)}_{l}=\left( 1-\hat{P}_{n} \right) [\phi^{(n)}_{l}, {\cal H}] \]
 with the Hamiltonian of the system ${\cal H}$ and commutator $[\cdot,\cdot]$.
Then the Green's function is obtained in a continued fraction form as
\begin{equation}
G_{k} (i\omega_{n}) = \dfrac{1}
                       {i\omega_{n}+\mu-\epsilon^{(1,1)}_{k} 
                            -\dfrac{\epsilon^{(2,1)}_{k}}
                             {i\omega_{n}+\mu- \epsilon^{(2,2)}_{k} 
                             -\Sigma^{(2)}(i\omega_{n})}
                             }.
\label{eqn:Gkw}
\end{equation}
where $\mu$ is the chemical potential and the matrices $\epsilon^{(1,1)}_{k\sigma}$, $\epsilon^{(2,1)}_{k\sigma}$, $\epsilon^{(2,2)}_{k\sigma}$ for the Hubbard model are given as~\cite{OnodaJPSJ,OnodaPRB}
\begin{subequations}
\begin{align}
\epsilon^{(1,1)}_{k \sigma}&=U \langle n_{-\sigma} \rangle - t_{k},\\
\epsilon^{(2,1)}_{k \sigma}&=U^{2}\langle n_{-\sigma} \rangle (1-\langle n_{-\sigma}\rangle ),\\
\epsilon^{(2,2)}_{k \sigma}&=U(1-\langle n_{-\sigma}\rangle )-\tilde{t}_{k}-\delta \mu,
\end{align}
\end{subequations}
where 
$t_{k}$ is the Fourier transformation of $t_{ij}$, which is defined as
\[
 t_{ij} =\left \{
 \begin{array}{ll}
 \displaystyle{ t } & \qquad (i,j):\ {\rm nearest\ neighbor} \\
 \displaystyle{ t^{\prime} } & \qquad (i,j):\ {\rm next\ nearest\ neighbor} \\
 \displaystyle{ 0 } & \qquad (i,j):\ {\rm otherwise } 
 \end{array} \right. 
 \]
 We have also introduced a related quantity 
$\tilde{t}_{k}$, which is the Fourier transformation of $\tilde{t}_{ij}$ defined as
\[ \tilde{t}_{ij}=\langle t_{ij}(\mib{S}_{i} \cdot \mib{S}_{j}+\frac{1}{4}n_{i}n_{j} -\Delta_{i}^{\dagger} \Delta_{j} ) \left(\langle n_{-\sigma} \rangle (1-\langle n_{-\sigma}\rangle )\right)^{-1},\]
where $\mib{S}_{i}$ and $\Delta_{i}$ are local spin and pair operators. These are defined as,
\[\mib{S}_{i}\equiv c^{\dagger}_{i\alpha}\mib \sigma_{\alpha \beta}c_{i\beta},\]with Pauli matrices $\mib \sigma$, and
\[\Delta_{i}\equiv c_{i\uparrow}c_{i\downarrow}, \]
respectively. 
The equal-time correlation functions of these quantities are calculated with the two-particle self-consistent method (TPSC)~\cite{TPSC}.
Finally, $\delta \mu$ is defined as
 \[ \delta \mu =-t_{ij} \langle c^{\dagger}_{i-\sigma}c_{j-\sigma} (1- 2n_{i\sigma})\rangle \left(\langle n_{-\sigma} \rangle (1-\langle n_{-\sigma}\rangle )\right)^{-1}. \]
 Through these equations, we explicitly included spin indices $\sigma$'s though we hereafter consider the spin-symmetric case.  
Next, in order to evaluate the quantity $\Sigma^{(2)}$, we introduce DMFT formulation~\cite{OnodaPRB}. Weiss fields for operators $c_{k\sigma}$ and $\phi^{(2)}_{k \sigma}$ are introduced as 
\begin{equation}
{\cal G}^{(0)}_{k}(i\omega_{n})=\dfrac{ 1 }{ i\omega_{n} +\mu-\epsilon^{(1,1)}_{k} - \epsilon^{(2,1)}{\cal G}^{(1)}(i\omega_{n}) }
\label{eqn:g0kw}
\end{equation}
\begin{equation}
{\cal G}^{(1)}(i\omega_{n})=( G_{\rm loc}^{(1)-1}(i\omega_{n})+\Sigma^{(2)}(i\omega_{n};\mu) )^{-1}.
\label{eqn:g1w}
\end{equation}
Here, $G_{\rm loc}^{(1)}$ is the local Green's function of the field $\phi^{(1)}$, which is formally given below. Then $\Sigma^{(2)}$ is obtained by a generalized form of the iterated perturbation theory (IPT) of the dynamical mean-field theory\cite{KotliarIPT} as,
\begin{equation}
\Sigma^{(2)}(i\omega_{n})=
\int_{0}^{\beta} d\tau e^{-i\omega_{n}\tau}\dfrac{1}{N^{3}} 
	\sum_{k,k^{\prime},q} 
		\dfrac{1}{\epsilon^{(2,1)}_{k}} \Gamma_{\mbox{$k,k^{\prime},q$}}
			{\cal G}^{(0)}_{\mbox{$k^{\prime}$}}(-\tau){\cal G}^{(0)}_{\mbox{$k^{\prime}+q$}}(\tau) {\cal G}^{(0)}_{\mbox{$k-q$}}(\tau),
\label{eqn:IPT1}
\end{equation}
where ${\cal G}^{(0)}_{k}$ is obtained self-consistently from Eqs.~(\ref{g0kw}),(\ref{g1w}) and 
\begin{equation}
G^{(1)}_{\rm loc}(i\omega_{n}) = \dfrac{1}{N} \sum_{k}  \dfrac{1}{i\omega_{n}+\mu-\epsilon_{k} ^{(2,2)}-\Sigma^{(2)}(i\omega_{n})},
\label{eqn:IPT2}
\end{equation}
Eqs. (\ref{eqn:g0kw}) to (\ref{eqn:IPT2}) constitute a set of self-consistent equations and are solved iteratively until the convergence.

CPM based on the original formulation shown above has achieved a certain success for bandwidth-control (BC) problems~\cite{OnodaPRB}. For filling-control (FC) problems, however, it shows some unphysical results. Namely, the filling changes even when the Fermi level is inside the gap at $T=0$, while in practice, it should not change within the gap. This failure comes from the careless treatment of the chemical potential as we explain in detail below. 

\subsection{Reformulation for FC problems}
 Now we propose a reformulation devised for FC problems.
\subsubsection{control over the implicit dependence of $\mu$}
In CPM Green's function of the form Eq.(\ref{eqn:Gkw}), its $\mu$ dependences not only come from explicit $\mu$, but also from implicit $\mu$ dependence of $\Sigma^{(2)}(i\omega_{n})$.
In order to control $\mu$, all the $\mu$-dependent parts have to be controlled consistently.
For implicitly $\mu$-dependent function $\Sigma^{(2)}(i\omega_{n})$, this is realized by recalculating it with the spectral representation of Green's function. Here, we show this procedure for ${\cal G}^{(0)}_{k} (i\omega_{n})$.

The procedure to determine the chemical potential is the following:

\noindent{1.} Calculate the spectral function ${\cal A}_{k}^{(0)}(\epsilon)$ of ${\cal G}^{(0)}_{k}(i\omega_{n})$ via Pad\'{e} approximation. 
Here, the spectral function is defined as \( {\cal A}_{k}^{(0)}(\epsilon)\equiv -\frac{1}{\pi}{\rm Im}{\cal G}^{(0)}_{k}(\epsilon+i\eta) \), where $\eta$ is a small positive number. 

\noindent{2. Re-calculate} the function ${\cal G}^{(0)}_{k}$ on the Matsubara axis with desired value of $\mu$ by spectral representation as 
\[\int d\epsilon {\cal A}^{(0)}_{k}(\epsilon)\dfrac{1}{i\omega_{n}+\mu-\epsilon}. \]
Note that in actual calculation, original ${\cal G}^{(0)}_{k}$ to be re-calculated is obtained with a certain value of $\mu|_{\scriptsize {\rm OLD}}$, which is represented as ${\cal G}^{(0)}_{k}(i\omega_{n};\mu|_{\scriptsize {\rm OLD}}).$ Then the Pad\'{e} approximation gives spectral function with its origin at $\mu|_{\scriptsize {\rm OLD}}$ as ${\cal A}^{(0)}_{k}(\omega ; \mu|_{\scriptsize {\rm OLD}}) \equiv {\cal A}^{(0)}_{k}( \epsilon=\omega+\mu|_{\scriptsize {\rm OLD}} )$. The new ${\cal G}^{(0)}_{k}$ with chemical potential $\mu$ is then obtained as 
\[{\cal G}^{(0)}_{k}(i\omega_{n};\mu)=\int d\omega {\cal A}^{(0)}_{k}(\omega;\mu|_{\scriptsize {\rm OLD}})\dfrac{1}{i\omega_{n}+\mu-\mu|_{\scriptsize {\rm OLD}}-\omega}. \]
This ${\cal G}^{(0)}_{k}$ is inserted to Eq.(\ref{eqn:IPT1}).

\subsubsection{Chemical potential for Weiss fields}
In CPM, Weiss fields for operator $c_{k\sigma}$ is given as Eq.(\ref{eqn:g0kw}). Here, the chemical potential $\mu$ appears in the manner just analogous to the lattice Green's function Eq. (\ref{eqn:Gkw}). In fact, however, this form is not obtained by equation of motion analysis like that in the lattice Green's function. Therefore it do not have to be necessarily equal to the chemical potential $\mu$ in the lattice Green's function. Here, in general, Weiss fields are represented with quantity $\tilde{\mu}$ as,
\begin{equation}
{\cal G}^{(0)}_{k}(i\omega_{n};\tilde{\mu})=\dfrac{ 1 }{ i\omega_{n}+\tilde{\mu}-\epsilon^{(1,1)}_{k} - \epsilon^{(2,1)}{\cal G}^{(1)}(i\omega_{n};\tilde{\mu}) }
\label{eqn:g0kwNEW}
\end{equation}
\begin{equation}
{\cal G}^{(1)}(i\omega_{n};\tilde{\mu})=( G_{\rm loc}^{(1)-1}(i\omega_{n};\tilde{\mu})+\Sigma^{(2)}(i\omega_{n};\tilde{\mu}) )^{-1}.
\label{eqn:g1wNEW}
\end{equation}
Here, in Eq.(\ref{eqn:g1wNEW}), $G_{\rm loc}^{(1)}(i\omega_{n};\tilde{\mu})$ is given by a similar form as the original CPM as \[G_{\rm loc}^{(1)}(i\omega_{n};\tilde{\mu})=\frac{1}{N}\sum_{k}\frac{1}{i\omega_{n}+\tilde{\mu}-\epsilon^{(2,2)}-\Sigma^{(2)}(i\omega_{n};\tilde{\mu})}. \]
In the original form of CPM, all $\tilde{\mu}$ is set equal to the chemical potential $\mu$ of the lattice Green's function. Here, we propose fixing it in a different way in order to satisfy other physical requirements.
We fix it so as to make the trace of the Weiss field ${\cal G}^{(0)}_{k}(i\omega_{n};\tilde{\mu})$ equal to that of the lattice Green's function $G_{k}(i\omega_{n};\mu)$, or equivalently, particle number $N_{e}$;

\[
\sum_{k}\sum_{n} {\cal G}^{(0)}_{k}(i\omega_{n};\tilde{\mu})=\sum_{k}\sum_{n} G_{k}(i\omega_{n};\mu) 
\]
The reason for adopting this condition will be given later.  
\subsubsection{Properties of the new formulation}
By adding these two reformulation, an important requirement for $T=0$ limit is satisfied.

\noindent{\it Requirement}:\\ 
If the spectral function is gapped, the result of calculation should not be affected by an arbitrary shift of chemical potential within the gap.

\noindent{\it Proof of the statement that the reformulated CPM satisfies this requirement}:\\
By the first reformulation, chemical potential is completely controlled and the following obvious relation is satisfied;
\begin{equation}
\frac{1}{\beta}\sum_{n} G_{k}(i\omega_{n};\mu)=\int d\epsilon A_{k}(\epsilon)f(\epsilon-\mu).
\label{eqn:prv1}
\end{equation}
Here, $A_{k}(\epsilon)$ is the spectral function of $G_{k}(i\omega_{n})$ defined as \( A_{k}(\epsilon)\equiv -\frac{1}{\pi}{\rm Im}G_{k}(\epsilon+i0) \), and function $f(\omega)$ is the usual Fermi distribution function.
In case of $T\to 0$ limit, Fermi distribution function becomes a step function. Then the right hand side (RHS) of Eq.(\ref{eqn:prv1}) becomes
\begin{equation}
({\rm RHS})=\int_{-\infty}^{\mu} d\epsilon A_{k}(\epsilon).
\end{equation}
and it is not affected by the shift of $\mu$ to $\mu^{\prime}$ as long as both of them are within the gap, {\it i.e.} 
\[\int_{-\infty}^{\mu^{\prime}} d\epsilon A_{k}(\epsilon)=\int_{-\infty}^{\mu} d\epsilon A_{k}(\epsilon) \] if \[\int_{\mu}^{\mu^{\prime}} d\epsilon A_{k}(\epsilon)=0. \]
Therefore, all the momentum distribution $\langle n_{k}\rangle$ as well as the total particle number $N_{e}$ is unaffected by the shift of chemical potential within the gap.
Next, by the second reformulation, if the particle number does not change, all Weiss fields are unchanged.
This is because $\tilde{\mu}$ as well as the matrix elements does not change.
Note that all the matrix elements such as $\epsilon^{(1,1)}$ are only dependent on $\mu$ and several lattice quantities such as \( \sum_{j}\langle t_{ij}c_{i\sigma}c_{j\sigma}\rangle=\frac{1}{N_{s}}\sum_{k}t_{k}\langle n_{k} \rangle \) which are unchanged here. 
Thus the whole results are unaffected.
Here, we note that the Weiss field is not necessarily gapped when the lattice Green's function has a gap. Therefore the result may be changed if one simply set $\tilde{\mu}$ equal to $\mu$. 
The idea of fixing the quantity $\tilde{\mu}$ different from $\mu$ to satisfy some physical requirement shares some similarity to the method proposed by Kajueter and Kotliar.~\cite{KajueterKotliar} They proposed a method of applying the iterated pertubation theory (IPT) for the dynamical mean-field theory (DMFT) to particle-hole asymmetric models~\cite{ref:KKipt}. Since CPM is an extention of DMFT with IPT, it may also suffer from the problem arising from applying IPT in particle-hole asymmetric models. The present prescription solves this difficulty.  Different solver of CPM could make this procedure unnecessary.

\section{Numerical Results}
In this section, we show the results of our numerical calculations. The following calculations have been performed on the $32\times32$ lattice with 1024 ($T=0.05$) or 512 ($T=0.10$) Matsubara frequencies.
 The spectral functions have been obtained by the analytic continuation using Pad\'{e} approximation.
 Throughout this paper the energy unit is set as $t=1$. 

\subsection{Phase diagram}
\begin{figure}
\begin{center}
\includegraphics[scale=0.70, angle=-90]{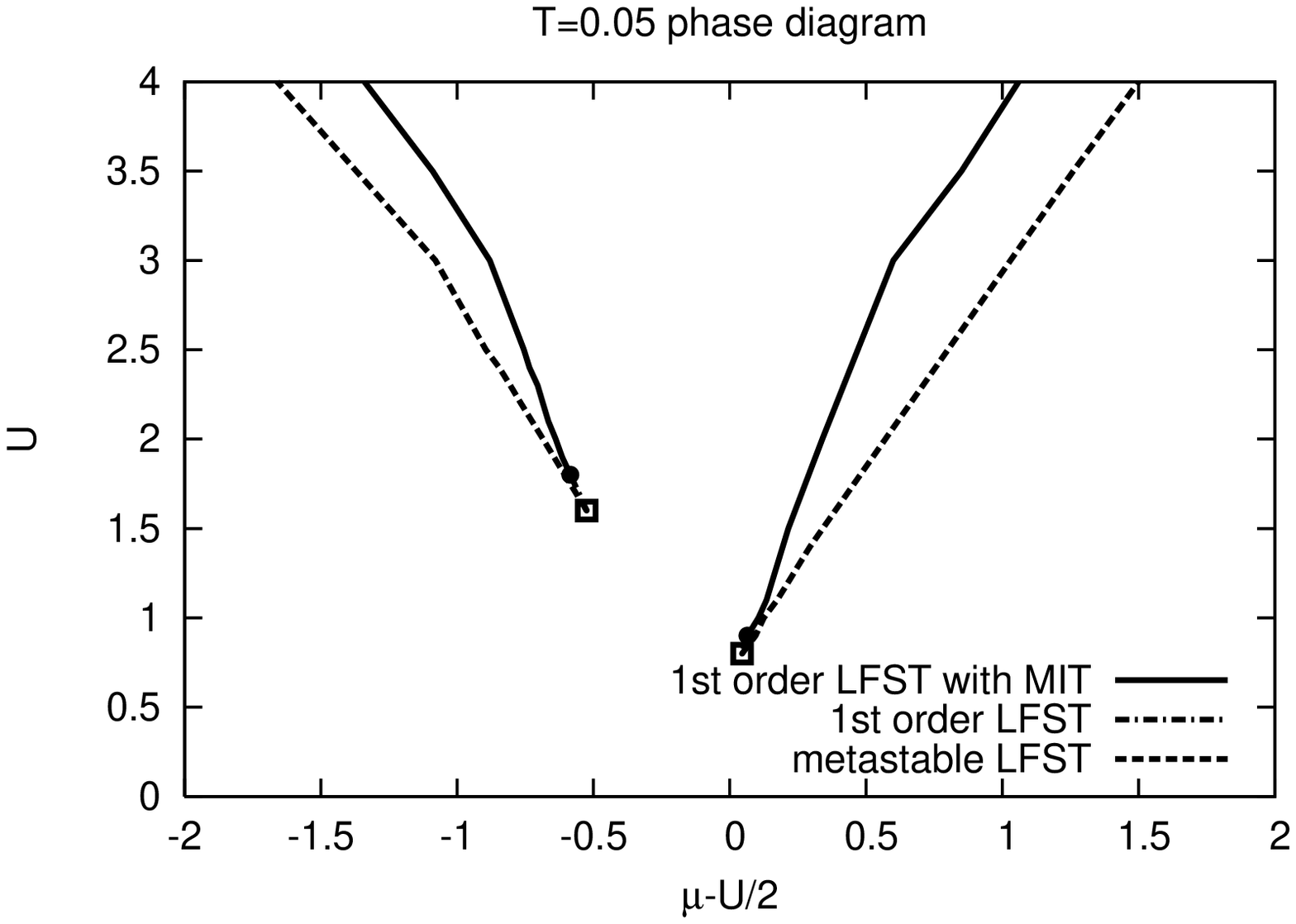}
\includegraphics[scale=0.70, angle=-90]{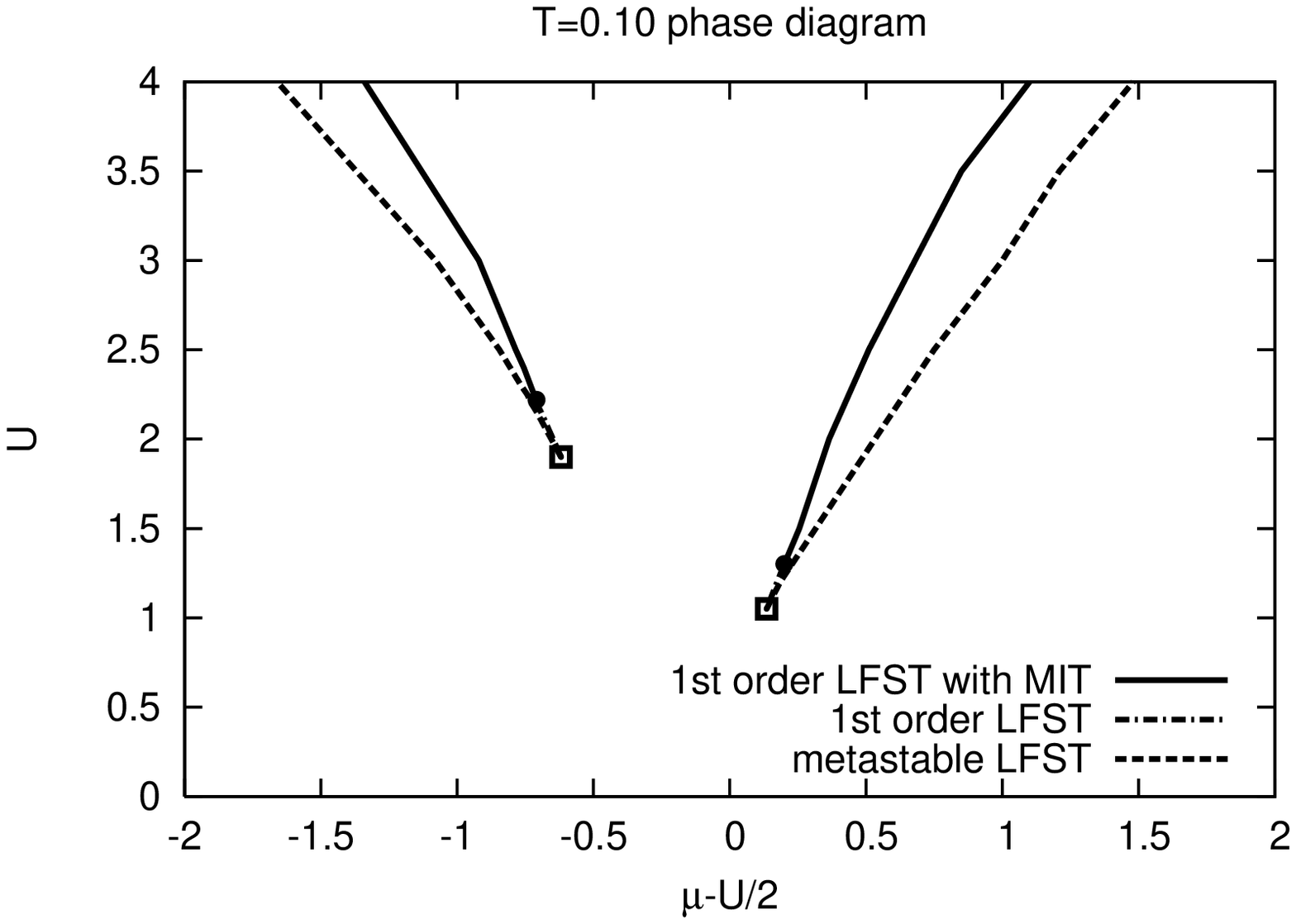}
\end{center}
\caption{
Phase diagram of the 2D Hubbard model at $t^{\prime}=-0.20$ in the plane of $\mu$ and $U$ at $T=0.05$ (upper panel) and at $T=0.1$ (lower panel). Solid lines indicate the simultaneous first-order Lifshitz transition and metal-insulator transition (MIT), while the dashed lines indicate the parameters with which the metastable solution becomes unstable. Metastable insulating solution exist in the regions sandwiched by the solid and the dashed lines.  The squares are the critical points of the Lifshitz transitions, or equivalently, the endpoints of the first-order Lifshitz transitions.
Note that the dash-dotted lines are drawn between the square and the circle in small segments of first-order Lifshitz-transition line and indicate first-order Lifshitz transitions without MIT. See Fig.\ref{fig:muUphaseL} for more detail. Circles, which separate solid lines and the dash-dotted lines are also explained in  Fig.\ref{fig:muUphaseL}.
} 
\label{fig:muUphase}
\end{figure}

\begin{figure}
\begin{center}
\includegraphics[scale=0.8, angle=-90]{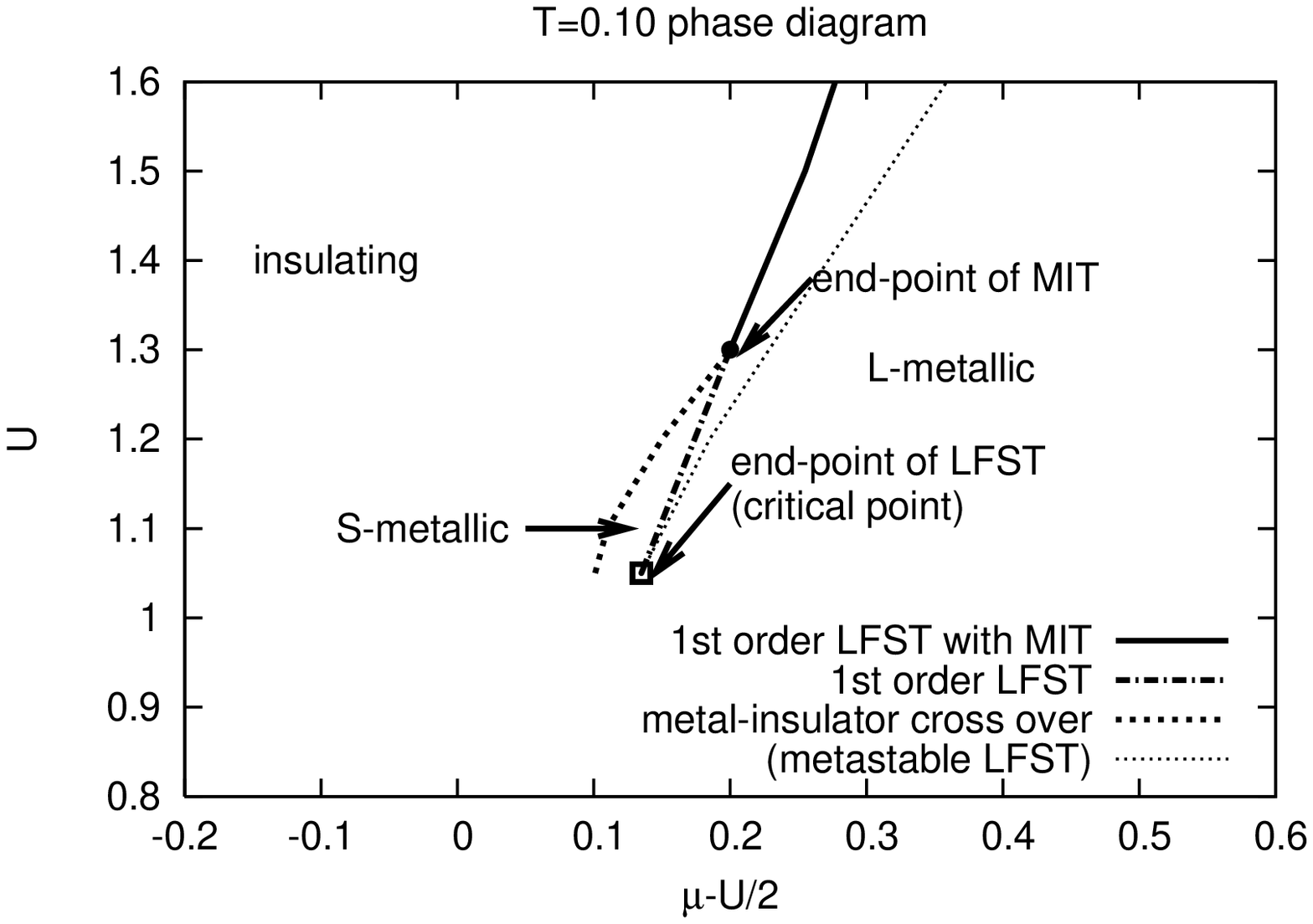}
\end{center}
\caption{
Detailed view of the $\mu$-$U$ phase diagram at $T=0.10$, enlarged from Fig.~\ref{fig:muUphase}.
Here, as the same as Fig. \ref{fig:muUphase}, the solid and dash-dotted lines indicate the first-order Lifshitz transition, where the solid line accompanies MIT while the dash-dotted line does not.
Thick dotted line indicates metal-insulator crossover lines. On this line, the bottom of the upper Hubbard band (UHB) crosses the Fermi level and the system crosses over between metallic phase and insulating phase. 
Note that such crossing of QP dispersion does not indicate a transition since the system is at a finite temperature. On the other hand, along the solid lines above the circles, the crossing occur discontinuously and one can define metal-insulator transitions.
The thin dashed lines indicate the parameters with which the metastable insulator solution becomes unstable.  
}
\label{fig:muUphaseL}
\end{figure}

\begin{figure}[t]
\begin{center}
\includegraphics[scale=0.55, angle=-90]{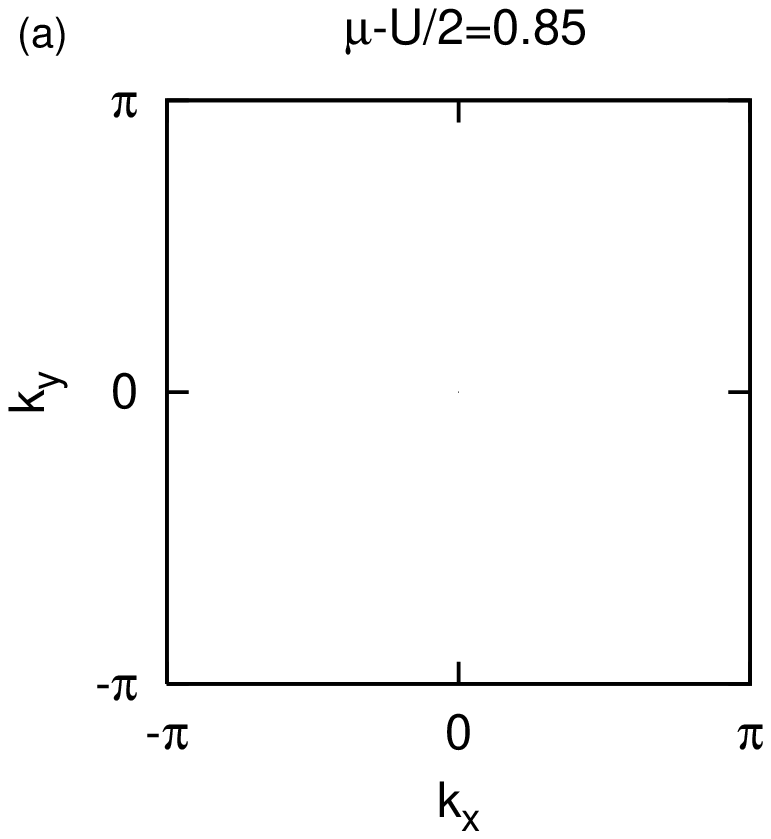}
\includegraphics[scale=0.55, angle=-90]{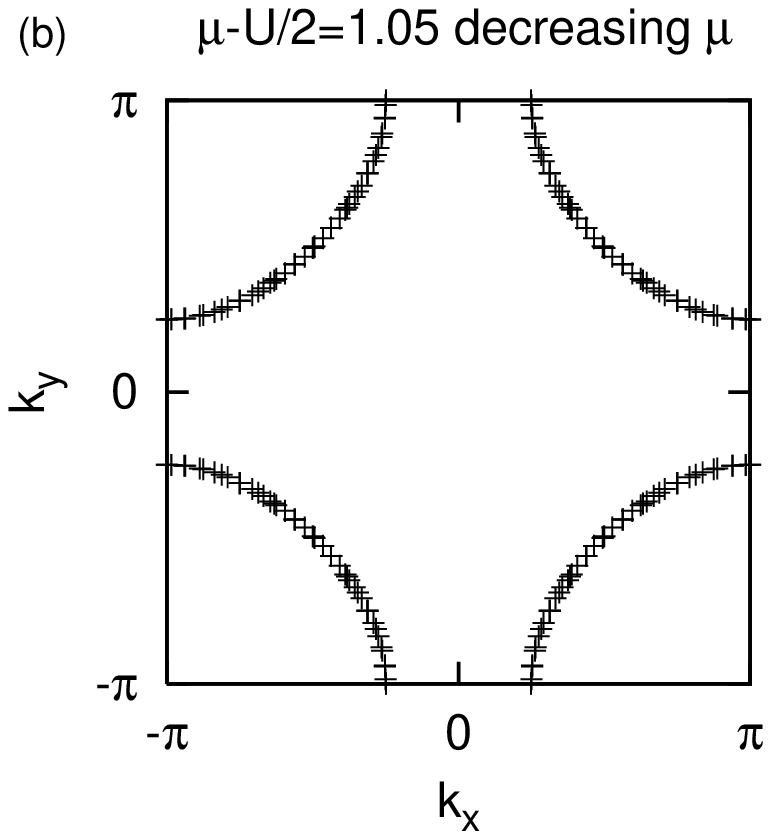}
\includegraphics[scale=0.55, angle=-90]{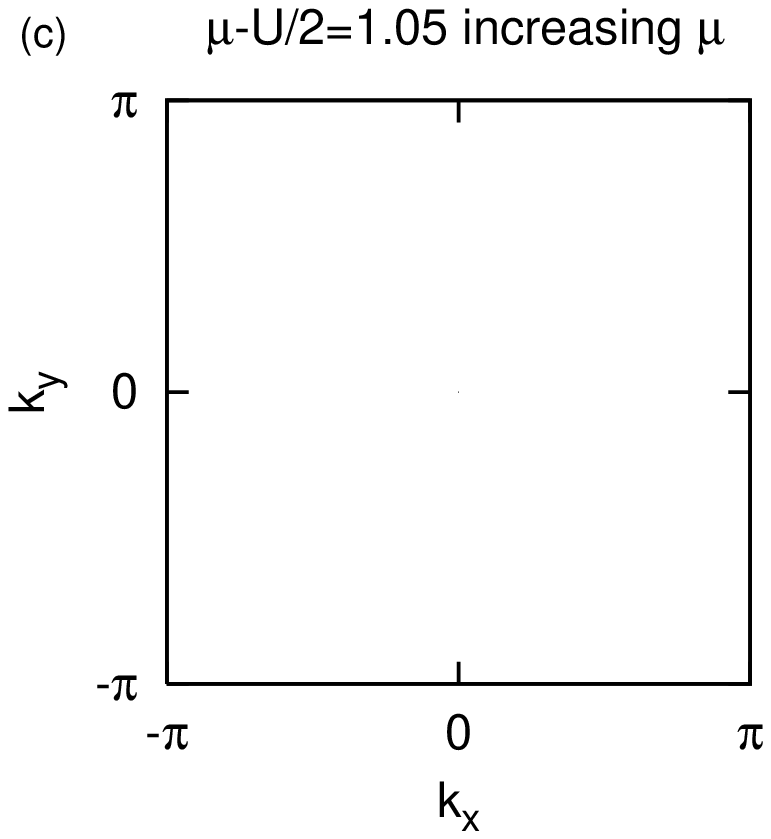}
\includegraphics[scale=0.55, angle=-90]{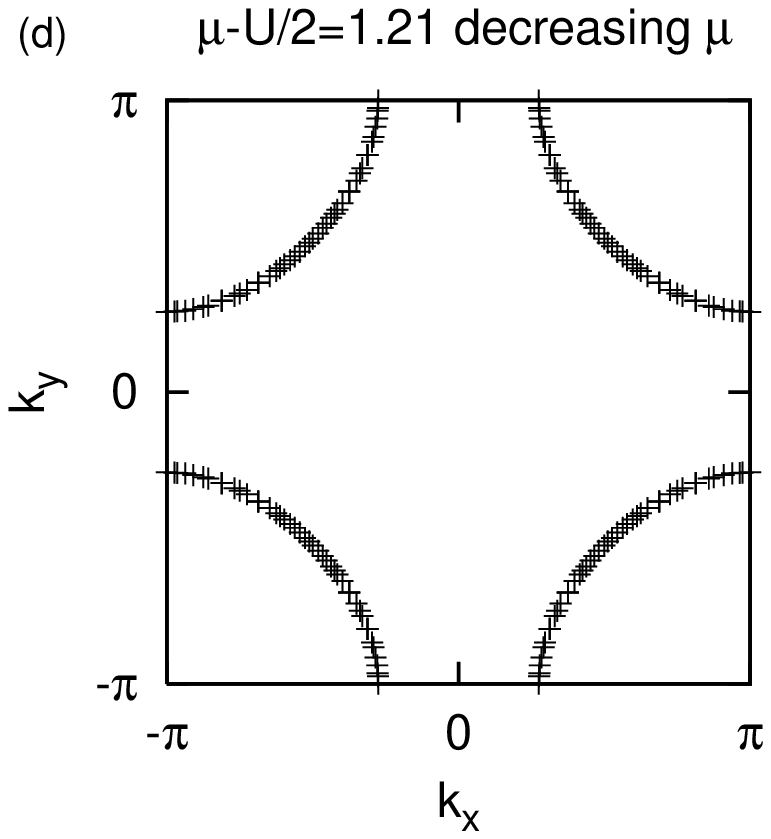}
\includegraphics[scale=0.55, angle=-90]{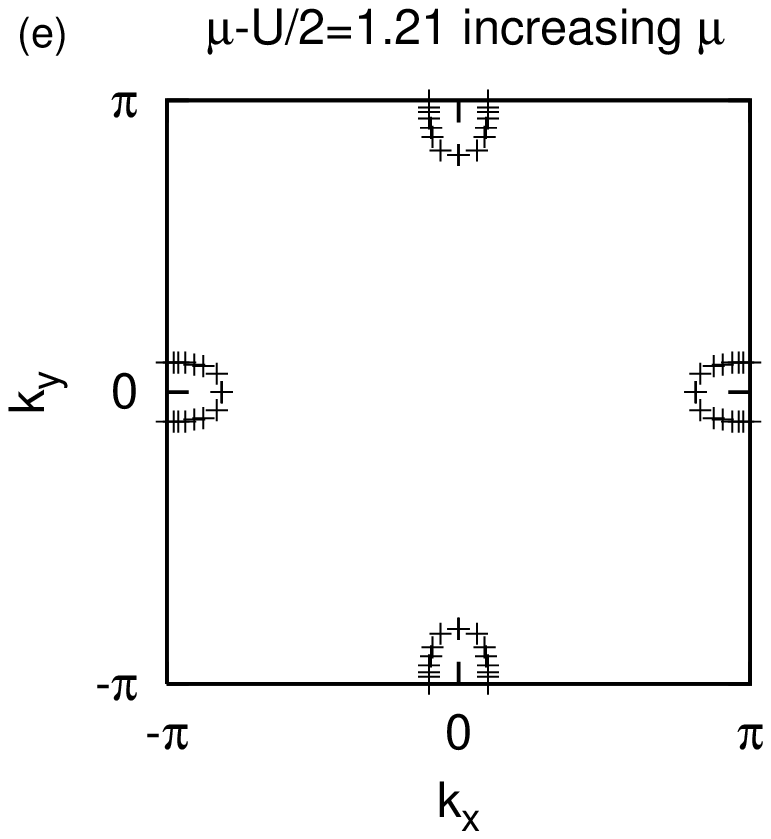}
\includegraphics[scale=0.55, angle=-90]{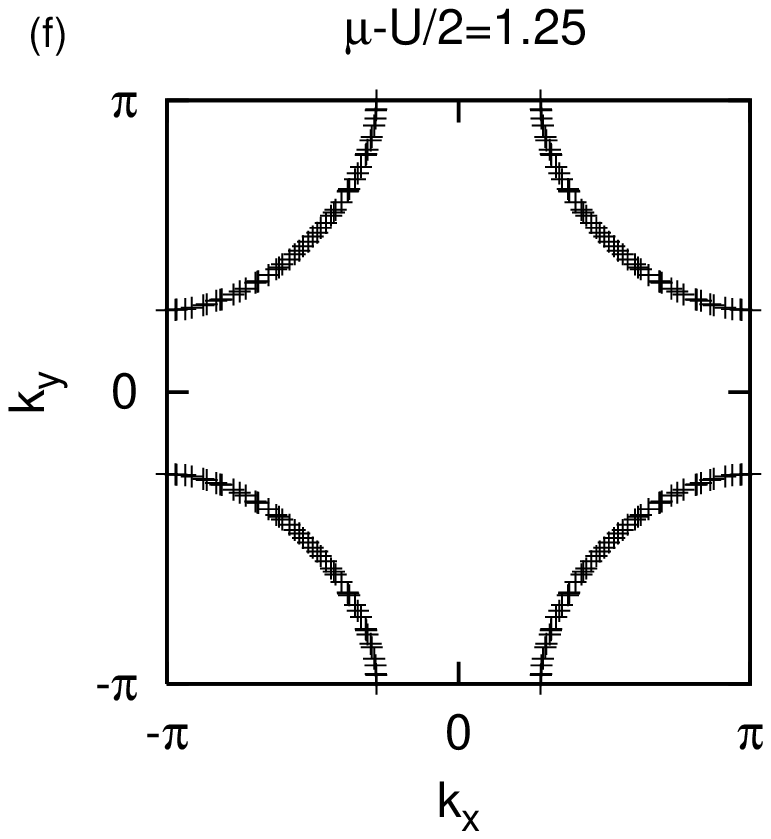}
\end{center}
\caption{
Fermi surfaces of solutions at fixed $U=3.50$ and different values of $\mu-U/2$. Starting from (f), in the sweep of decreasing $\mu$, we obtain (f)-(d)-(b)-(a) in a sequence, while in the sweep of increasing $\mu$, we obtain (a)-(c)-(e)-(f).
Here, at $\mu-U/2=0.85$ and at $\mu-U/2=1.05$ of the $\mu$-increasing sweep, the solutions are insulating and have no Fermi surface.
There is a discontinuous transition between (b) and (a) in the former sweep, while between (e) and (f) in the latter sweep.
The transition (e) to (f) is not a metal-insulator transition, but a Lifshitz transition from the small Fermi surface (S-phase) to the large connected Fermi surface (L-phase). At the transition (b) to (a), the Lifshitz transition accompany a metal-insulator transition. 
As is mentioned in the text, the solutions obtained through decreasing $\mu$ are the true equilibrium
}
\label{fig:FSplotT010u35}
\end{figure}

First, we show phase diagram of the Hubbard model at fixed temperatures. At $T=0.05$ and $0.10$, the $\mu$-$U$ phase diagram is obtained as Fig. \ref{fig:muUphase} (For an enlarged figure see also Fig.\ref{fig:muUphaseL}) . Here, first-order transitions appear with a certain range of coexsistence of two solutions. 
The phase transition obtained here contains Lifshitz transitions, which is the transition caused by the topology change of the Fermi surface. In Fig.\ref{fig:muUphase}, roughly speaking, the equilibrium solution shows a first-order transition from metals with a large Fermi surface to insulators.  However, if one carefully examines the metastable solutions obtained from the adiabatic continuation of the insulating phase, it turns out that the Fermi surface changes from large to small or vice versa as is shown in Fig. \ref{fig:FSplotT010u35}. Therefore, it turns out that the Lifshitz transition occurs simultaneously with the metal-insulator transition. The small Fermi surface is preempted in the equilibrium because of the first-order transition.
We refer to the phase with the small Fermi surface as S-phase and that with the large Fermi surface as L-phase. We also use the expression such as S-metallic phase and L-metallic phase in order to describe metallic phase with small Fermi suface and that with large Fermi surface, respectively.  Although the S-phase in Fig.\ref{fig:FSplotT010u35} stays as a metastable one, we will show below that the S-phase indeed becomes the equilibrium in the region between the square and the filled circle in Fig.\ref{fig:muUphase}.

We note that Lifshitz transitions as well as metal-insulator transitions are originally transitions in the zero-temperature limit. However, they are extended to finite temperature if these transitions are of the first order.
Finite temperature Lifshitz (metal-insulator) transitions are defined as the first-order transitions from phase $A$ to phase $B$ where the zero-temperature extention of the phase $A$ and the phase $B$ are separated by the Lifshitz (metal-insulator) transitions of the original definitions.
The spectral density distribution changes discontinuously at first-order Lifshitz transitions. 
Therefore they accompany metal-insulator transitions if the spectral density changes from metallic to insulating. 
In the phase diagram in Fig.~\ref{fig:muUphase} as well as in Fig.~\ref{fig:muUphaseL}, solid lines above the circles show such kind of transitions. 

\begin{figure}
\begin{center}
\includegraphics[scale=0.8, angle=-90]{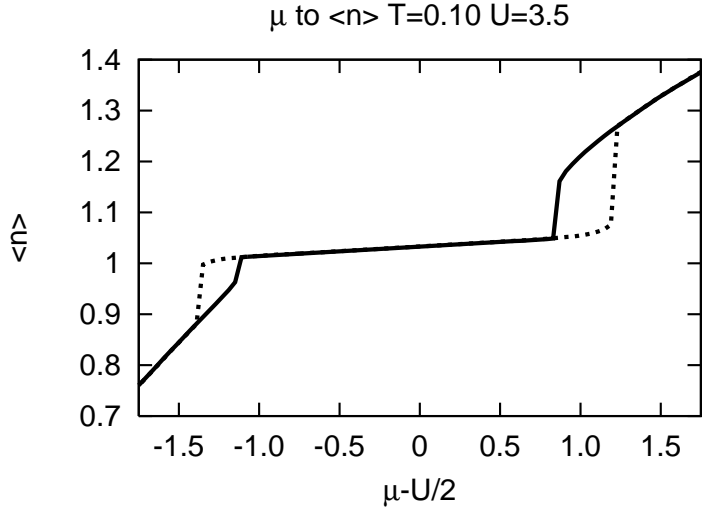}
\end{center}
\caption{
Plot of filling $\langle n \rangle$ at $T=0.10$, $U=3.50$ as a function of the chemical potential $\mu$. The solid line with $\mu-U/2<0$ ($\mu-U/2>0$) was obtained by sweeping $\mu$ from the lowest (highest) values of $\mu$ to $\mu-U/2=0$, while the dashed line was obtained by sweeping $\mu$ in the opposite direction.
Note that the full line corresponds to the sweep from the metallic to the insulating phase, and the dashed line, opposite. 
}
\label{fig:nmuplotT010u35}
\end{figure}

\begin{figure}
\begin{center}
\includegraphics[scale=0.8, angle=-90]{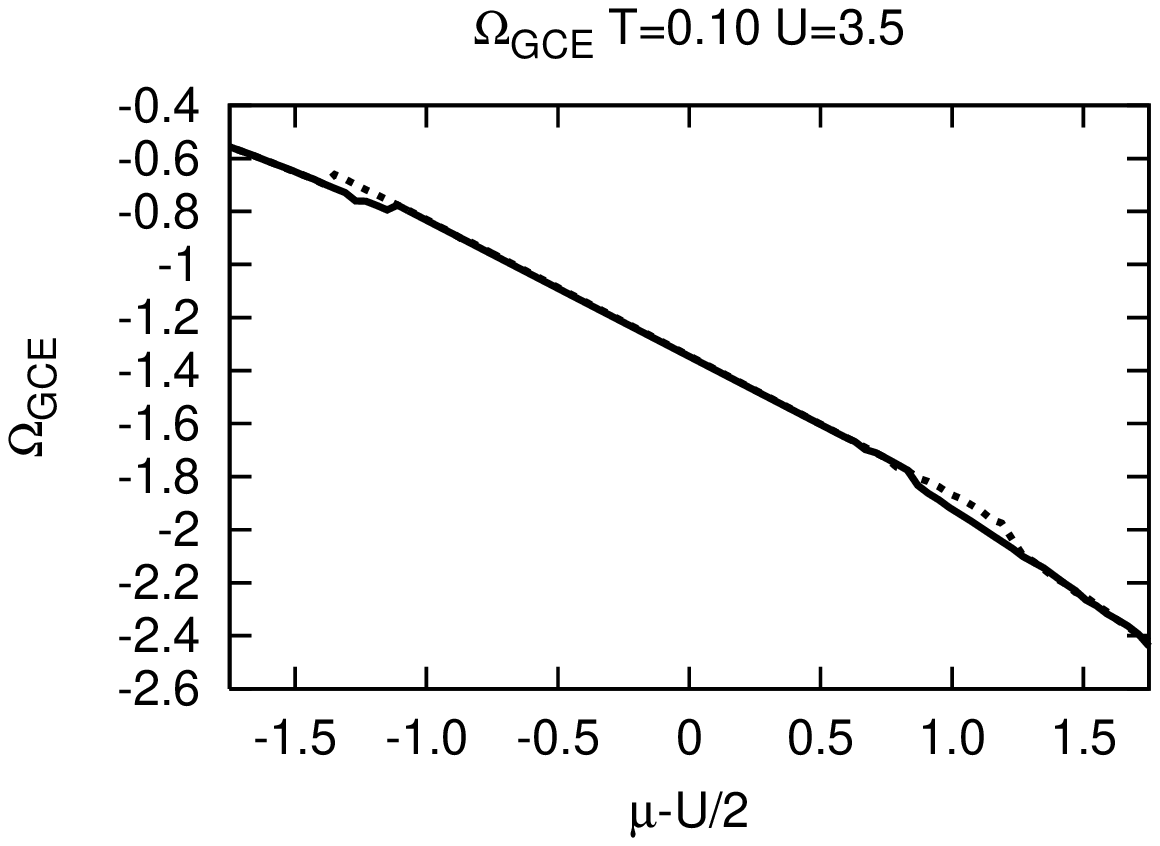}
\end{center}
\caption{ 
Plot of the grand canonical potential $\Omega_{\rm GCE}$ for the solutions shown in Fig. \ref{fig:nmuplotT010u35}. Solid and dashed lines show the sweep in the different directions as is shown in Fig. \ref{fig:nmuplotT010u35}. The result shows that the solutions along the solid line are the equilibrium ones within error bars.
}
\label{fig:muOMEGAplotT010u35}
\end{figure}

We consider the region above the solid circle in Fig.\ref{fig:muUphase} in more detail, where the first-order Lifshitz transition occurs simultaneously with MIT.  Typical calculated results for the density and the grand canonical potential at a fixed value of $U$, $U=3.50$ and sweeping $\mu$ in different directions are shown in Figs. \ref{fig:nmuplotT010u35} and \ref{fig:muOMEGAplotT010u35}. Here, a clear hysteresis is observed.
One series of solutions is obtained by sweeping $\mu$ from the lowest values of $\mu$ to $\mu=U/2$, while the other series of solutions is obtained by sweeping $\mu$ in the opposite direction.
As will be explained below, each corresponds to the sweep from the hole-doped (electron-doped) metallic phase to the insulating phase and vice versa, respectively.
To clarify the nature of each solution, we take the electron-doping side for example, and plotted the Fermi surface of each solution in Fig. \ref{fig:FSplotT010u35}. Here, each Fermi surface is obtained as the zero-energy section of the QP dispersion.
The solutions obtained through sweeping $\mu$ in the increasing order, which are metastable solutions as we will explain below, are solutions with the small Fermi surfaces ( S-phase ). Those obtained through sweeping $\mu$ in the decreasing order, which are equilibrium solutions, are those with larger Fermi surfaces ( L-phase ).
We compare the thermodynamic potential $\Omega|_{\rm GCE}=F-\mu \langle N_{e}\rangle$, where $F$ is the Helmholtz's free energy, for the both solutions. As shown in Fig. \ref{fig:muOMEGAplotT010u35}, the L-phase has lower potential compared to the S-phase in most part of the coexistence region, and the equilibrium phase boundary is very close to the point where the L-phase solution becomes unstable.

%
In this parameter region, spectral density of the metastable region of the S-phase changes from the metallic to insulating as the system approaches half filling. It is insulating at the point where the true phase transition occurs.
Therefore, the transition here accompanies a metal-insulator transition. 

\begin{figure}[h]
\begin{center}
\includegraphics[scale=0.8, angle=-90]{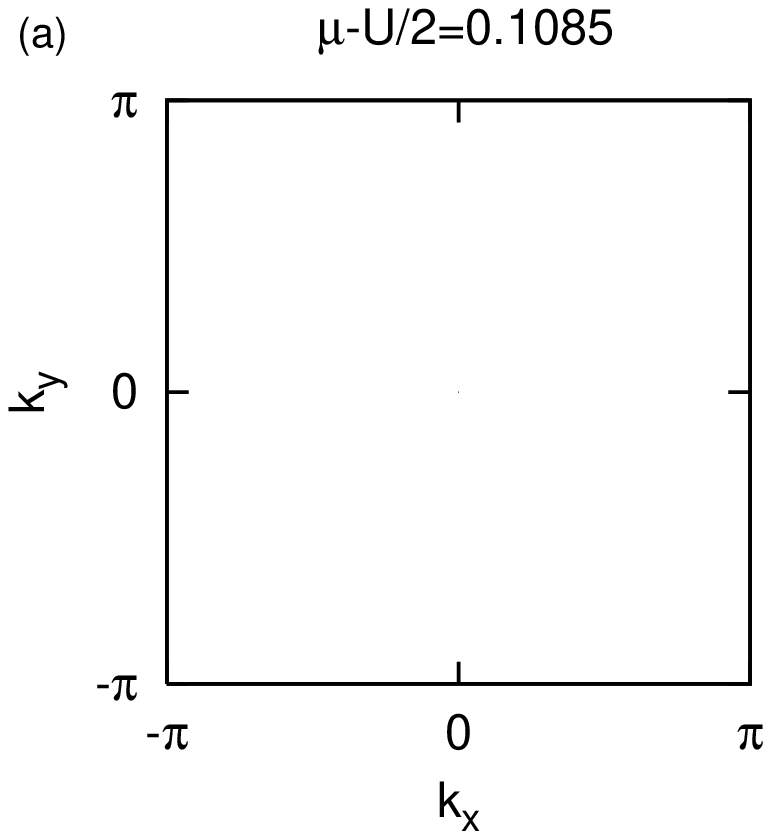}
\includegraphics[scale=0.8, angle=-90]{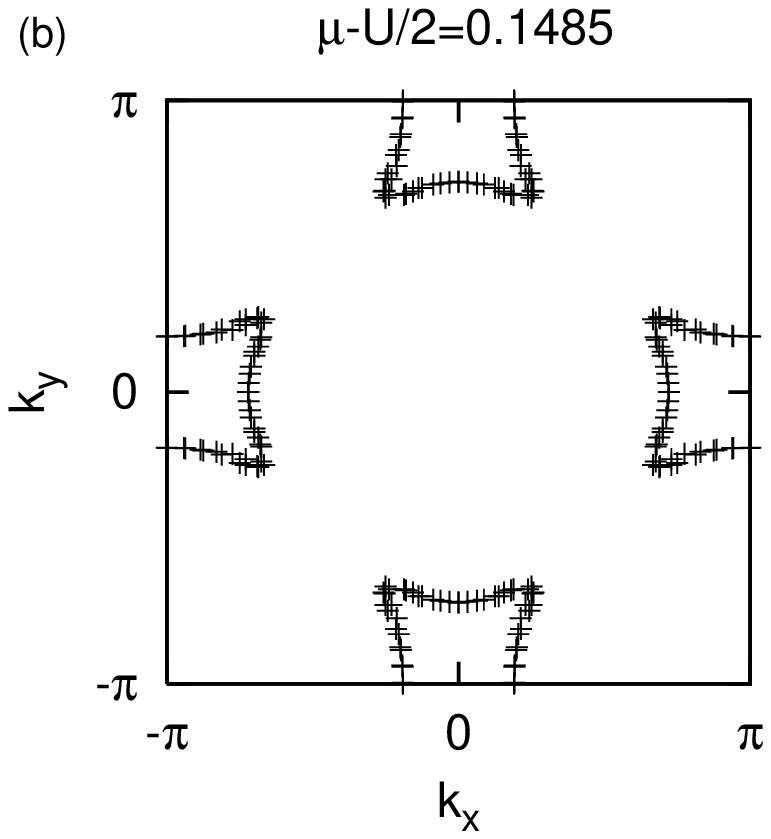}
\includegraphics[scale=0.8, angle=-90]{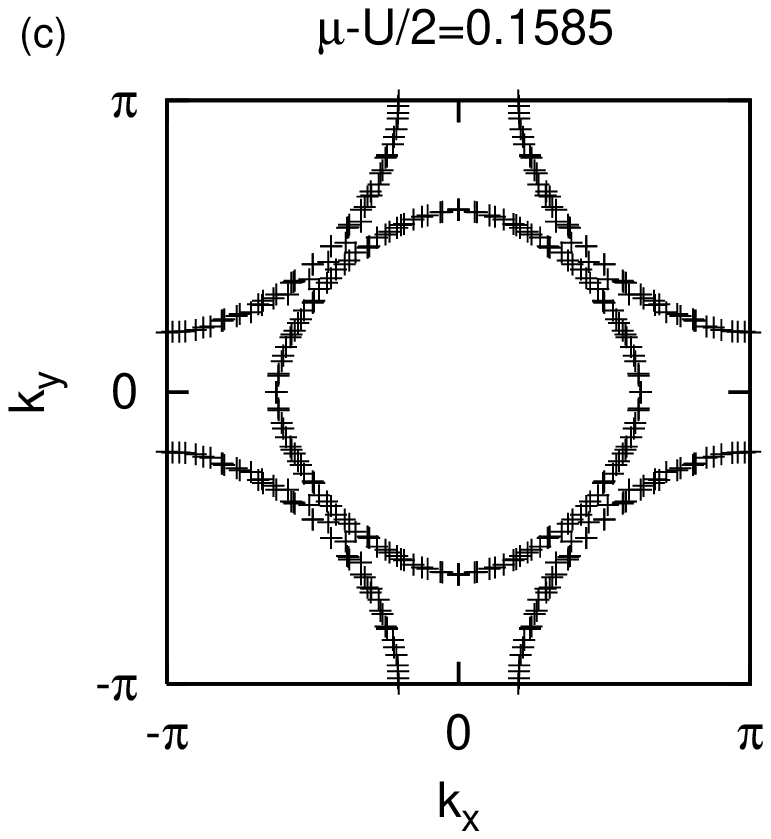}
\includegraphics[scale=0.8, angle=-90]{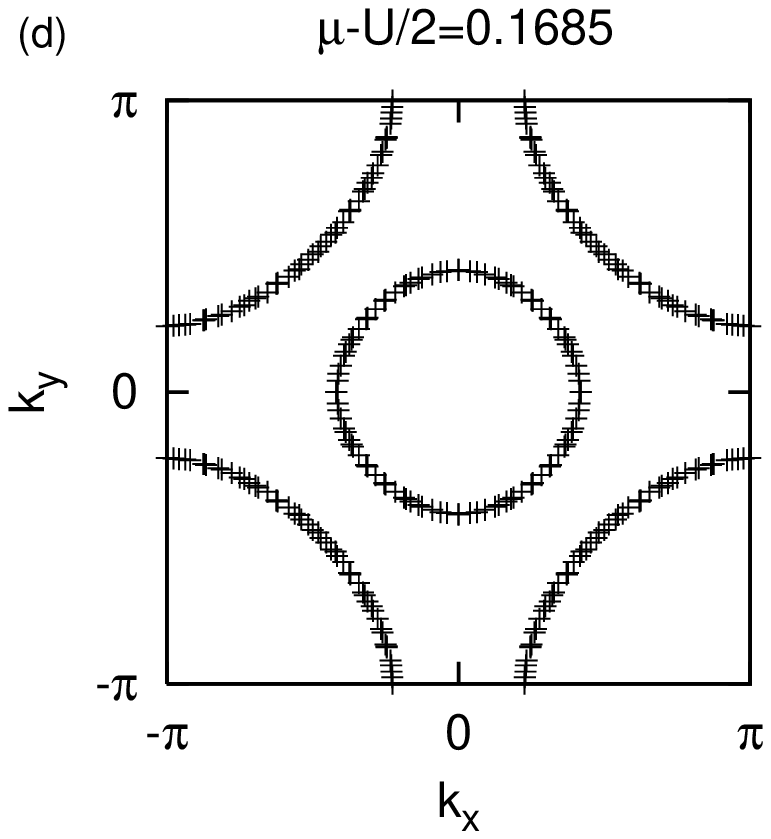}
\end{center}
\caption{
The Fermi surface around the critical point at fixed value of $U$; $U=U_{c}=1.05$, and different values of $\mu$; $\mu-U/2=0.1085$, $0.1485$, $0.1585$, and $0.1685$, respectively. 
Note that at $\mu-U/2=0.1085$, Fermi surface does not exist since it is in the insulating phase.
As the chemical potential increases from $\mu-U/2=0.1085$,
the system crosses over from insulating phase ($\mu-U/2=0.1085$) to metallic phase with small Fermi surfaces ( S-metallic ) ($\mu-U/2=0.1485$). It then crosses the critical point ($\mu-U/2=0.1585$), and finally metallic phase with large connected Fermi surface ( L-metallic ) ($\mu-U/2=0.1685$).  The weight of the inner circle in (c) and (d) is much smaller than that of outer open curves.  With further evolution of doping beyond (d), the inner circle in (d) fades out and disappears which leaves the single Fermi surface evolved from the outer curves.  
}
\label{fig:FSplotT010u1050}
\end{figure}

Now we discuss the detailed view of the $\mu$-$U$ phase diagram shown in Fig.\ref{fig:muUphaseL} in detail. Along the dash-dotted lines below the circles, first-order Lifshitz transition occurs from S-metallic to L-metallic and the metal-insulator transition becomes just a continuous crossover separated from the Lifshitz transition. 
Here, we investigate the behavior around the critical point at fixed $U=U_{c}$ for several choices of $\mu$.
As is shown in Fig.\ref{fig:muUphaseL}, as the chemical potential increases from the insulating region at $U=U_{c}$, the system crosses over from the insulating phase to the the S-metallic phase. Then it crosses the critical point, and finally, enters the L-metallic phase.  We have shown the evolution of the Fermi surface in Fig.\ref{fig:FSplotT010u1050}.
Within the rigid band picture, the insulating phase crosses over to the S-metallic phase as the chemical potential $\mu$ crosses the bottoms of the upper Hubbard band (UHB), which are the M-points (here defined as the momentum $(\pi,0)$ and its equivalent points). 
In the S-metallic phase, the section of the Fermi surface is disconnected pockets around the M points. It changes to connected large one at the Lifshitz transition point as the Fermi level crosses the X-points (here defined as the momentum $(\pi/2,\pi/2)$ and its equivalent points).
In this intermediate region, the S-metallic phase is indeed the equilibrium.  Apparently, the S-metallic phase does not satisfy the Luttinger sum rule~\cite{Luttinger}.  Since the Mott gap is formed without any long-ranged order in this formalism, the Luttinger volume cannot be satisfied near half filling unless this S-metallic phase is preempted by a strong first-order transition. The violation of Luttinger sum rule in underdoped region was also suggested in cluster DMFT~\cite{Jarrell}.  At the transition in the electron-doped side, appearance of Fermi pockets in the underdoped region similar to those in the S-metallic phase was shown in certain approximations~\cite{Kusunose,Kusko}. In their cases, however, the symmetry breaking by the antiferromagnetic order was assumed as the starting point, while we have not assumed the symmetry breaking in the present study. At nonzero temperatures, the S-metallic phase is in any case adiabatically connected with the insulating phase and a strict statement on the Luttinger sum rule is neither possible nor meaningful. Although it is remarkable that the apparent small Fermi pocket is detected in our study at low but finite temperatures, it does not necessarily exclude that Lifshitz transition and MIT always merge at strictly zero temperature, thereby the S-phase is preempted by the first-order transition.  However, the critical temperatures of Lifshitz and metal-insulator transitions may be lowered to zero by enhancing quantum fluctuations, where the first-order transitions disappear.  If the both transitions become continuous at zero temperature, we do not know how to make them compatible with the Luttinger sum rule, if the quasiparticle weight remains nonzero through the transitions as we clarify later.   The gap formation without the symmetry breaking shares a similarity to the renormalization group results~\cite{Rice}
The region of this separation between metal-insulator and Lifshitz transitions may be more extended for larger $t'$ with larger region of S-metallic phase.
 
The S-metallic phase sandwiched by the Lifshitz transition and the metal-insulator transition in the underdoped region may show unusual metallic behavior far from the simple Fermi liquid. Near the S-metallic phase, the small Fermi volume as in Fig.\ref{fig:FSplotT010u1050}(c) has very anisotropic Fermi surface and the inner circle has faint weight as compared to the outer surface.  When the faint inner surface becomes not observable in the experimental resolution, this may lead to a large Fermi volume obtained from the outer curve. The volume quickly approaches that expected from the Luttinger theorem with evolving doping.  At the same time, this anisotropy in the case of Fig.\ref{fig:FSplotT010u1050}(b) may lead to observation of an arc structure of the Fermi surface only consisting of a part of the Fermi surface.  In the hole doped case, this makes the arc structure around $(\pi/2,\pi/2)$ as is observed in angle resolved photoemission experiments of the cuprate superconductors\cite{Yoshida1,Yoshida2}.  The Luttinger sum rule may be approximately satisfied for relatively small $t'$, while it may be strongly violated in the S-phase as well as in the situation like Fig.\ref{fig:FSplotT010u1050}(c) with increasing $t'$ at a relatively small $U$ region.

\subsection{Determination of critical points}
\begin{figure}
\begin{center}
\includegraphics[scale=0.8, angle=-90]{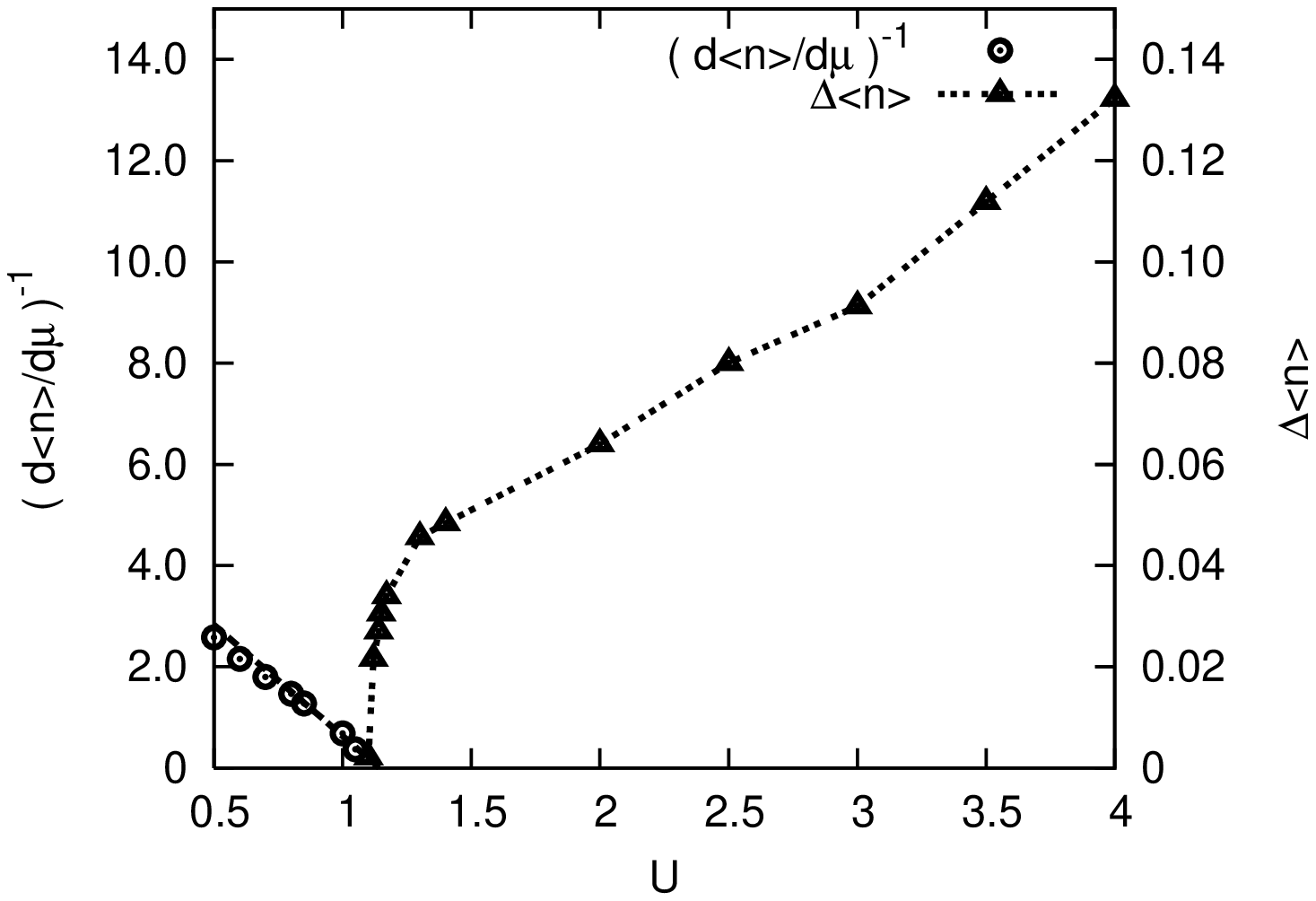}
\end{center}
\caption{
Singularities around the electron-like critical point. 
Data points are taken along the first-order Lifshitz-transition line and its extension, crossover line (see text).
From the continuous transition side, from lower $U$, minimum values of inverse charge susceptibility $(dn/d\mu)^{-1}$ are plotted with circles, while from the discontinuous transition side, jumps of the filling $\Delta n$ at the Lifshitz transition points are plotted with triangles.
Here, the circles are extrapolated linearly to the zero-point at $U=1.05$. 
The triangles do not show linear behavior, and are monotonic, which also decreases to zero near $U_c=1.05$. These suggest the exponents $\gamma=1$ and $\beta<1$, which are consistent with the mean-field Ising values $\gamma=1$ and $\beta=1/2$, although quantitative estimates are left for future studies.  Note that the exponents are defined by $(dn/d\mu)^{-1} \propto |U-U_c|^{\gamma}$ and $\Delta n \propto |U-U_c|^{\beta}$.
}
\label{fig:T010extrapE}
\end{figure}

As is shown in Fig. \ref{fig:muUphase} there are critical points at the ends of the first-order phase transition lines. The location of the critical points are carefully determined by an extrapolation procedure which is described below : 
We have obtained the first-order phase transition lines. 
Beyond the Lifshitz critical points, which are shown as squares in Fig. \ref{fig:muUphase}, there are crossover lines, though the crossover lines are not shown explicitly in the figure. 
The crossover lines are the extensions of the first-order phase transition lines and defined as the maximum of the charge susceptibility $\chi_{c}=\frac{ \partial <n> }{ \partial \mu }.$ 
The precise location of the critical points are found along this line. From the first-order transition side, it is found by extrapolating the first-order jump to zero. From the crossover side, it is found by extrapolating the inverse of the charge susceptibility to zero. An example of this extrapolation is shown in Fig. \ref{fig:T010extrapE}.
 
 \subsection{Analysis of the critical points}
 One of the major subjects of this research is to investigate the nature of these critical points. Now we analyze them in detail.
 The critical points are the points where the charge susceptibility $\chi_{c}$ diverges. The charge susceptibility is related to the spectral density function $A_{k}(\omega)$ as
 \begin{equation}
 \chi_{c}=\frac{\partial \langle n \rangle}{\partial \mu}
 =\frac{\partial}{\partial \mu} \int d^{2}k \int d\epsilon A_{k}(\epsilon) f(\epsilon-\mu)
 \label{eqn:chi_c1}
 \end{equation}
 Now we introduce the concept of quasiparticle (QP).
 We label each QP by its momentum $k$ and index $\alpha$, where the latter is introduced in order to distinguish different QP peaks with the same momentum $k$.
 The QP with label $k$, $\alpha$ is characterized by its pole energy $\epsilon_{k\alpha}$, and the pole weight $Z_{k\alpha}$. Then the spectral function $A_{k}(\omega)$ is devided into QP contribution and the remaining incoherent part as
 \begin{equation}
 A_{k}(\epsilon)=\sum_{\alpha}Z_{k\alpha}\delta(\epsilon-\epsilon_{k\alpha})+A_{k}^{\rm incoh}(\epsilon).
\label{eqn:Akwdevide}
 \end{equation}

After inserting Eq.(\ref{eqn:Akwdevide}) to Eq.(\ref{eqn:chi_c1}) we obtain the following expression for $\chi_{c}$ ;
 \begin{multline}
\chi_{c}=\frac{\partial}{\partial \mu} \int d^{2}k \sum_{\alpha}Z_{k\alpha}f(\epsilon_{k\alpha}-\mu)
+\frac{\partial}{\partial \mu} \int d^{2}k \int d\epsilon A_{k}^{\rm incoh}(\epsilon) f(\epsilon-\mu) \\
=\int d\epsilon^{\prime} \int_{\epsilon_{k}=\epsilon^{\prime}} dk \sum_{\alpha}\frac{Z_{k\alpha}}{|\nabla \epsilon_{k\alpha}|} \frac{df}{d\mu}(\epsilon^{\prime})
+\frac{d}{d\mu} \int d^{2}k \int d\epsilon A_{k}^{\rm incoh}(\epsilon) f(\epsilon-\mu). 
\label{eqn:chi_c2}
\end{multline}
If the concept of QP plays an important role, the first term in the last line of Eq. (\ref{eqn:chi_c2}) becomes dominant, and the divergence of the charge susceptibility occurs through the term $\frac{1}{|\nabla \epsilon_{k\alpha}|}$, {\it i.e.}, flattening of the QP dispersion. 

Next we show the procedure which we have taken in order to derive the above quantities from the numerical calculation.
 
 \noindent{1. Quasiparticle (QP) dispersion $\epsilon_{k\alpha}$ }\\
 We calculate spectral functions with small imaginary part $\eta$ as $A_{k}^{\eta}(\omega;\mu )\equiv -\frac{1}{\pi} {\rm Im}G_{k}(\omega+i\eta;\mu)$. QP peaks are defined as the peak structure in $A_{k}^{\eta}(\omega;\mu)$ with intensity larger than 1\% of its maximum value $\frac{1}{\pi\eta}$.

\noindent{2. Quasiparticle weights}\\
We define the QP pole weight $Z_{\rm QP}(k,\alpha)$ as an integration of spectral function around the QP pole;
\[ Z_{\rm QP}(k,\alpha) \equiv \int_{-\Delta/2}^{\Delta/2}d\omega^{\prime} A_{k}(\omega^{\prime}+\epsilon_{k\alpha} ). \]
Here, $\Delta$ is the width of energy window. We fix it as $\Delta=T$. This quasiparticle weight is computed for the pole, which can be adiabatically connected to the quasiparticle at the Fermi level in the metallic side.
 A related quantity is their Fermi-level contribution $Z_{\rm FS}(k)$, which is defined as an integration of spectral function around the Fermi level;
 \[ Z_{\rm FS}(k) \equiv \int_{-\Delta/2}^{\Delta/2}d\omega^{\prime} A_{k}(\omega^{\prime}). \]
 
\noindent{3. Effective Fermi momentum $\delta k_{F}^{\rm eff}(\delta \mu)$}\\
 Here, we define the Fermi surface by the zero energy section of the QP energy dispersion surface.   
 Around the critical point of the Lifshitz transition on the L-phase side, we introduce the following quantities to estimate the chemical potential dependence of the evolution of the Fermi surface:
 At the critical point, the Fermi surface has a topological singlularity. We define the momentum of this point as $k_{\rm F c}$.
 In the L-phase side, the Fermi surface expands around $k_{\rm F c}$. We measure the Fermi momentum at chemical potential $\mu$, $k_{F}(\mu)$ from the critical $k_{\rm F c}$. This quantity is a function of chemical potential measured from the critical chemical potential $\mu_{c}$.
 Now we define $\delta \mu\equiv |\mu-\mu_{c}|$, and  $\delta k_{F}^{\rm eff}(\delta \mu)\equiv k_{F}(\mu)-k_{\rm F c}.$
We note that in the rigid band limit, $\delta k_{F}^{\rm eff}(\delta \mu)$ is equivalent to the QP dispersion at the critical point, whereas it is not in case the QP dispersion changes with the chemical potential. 

Through the analysis, $xy$-symmetry is always conserved. Therefore, as for momentum space quantities, $xy$-symmetric points are equivalent. Hereafter, we refer to $(\pi,0)$ and its equivalent points in the momentum space as M-points, and  $(\pi/2,\pi/2)$ and its equivalent points in the momentum space as X-points. 
With these preparations, we analyze one of the obtained critical points. We choose the critical point at $T=0.10$ and study transitions to electron-like Fermi surface.

\subsection{QP dispersion around the critical point}
\begin{figure}
\begin{center}
\includegraphics[scale=0.60, angle=-90]{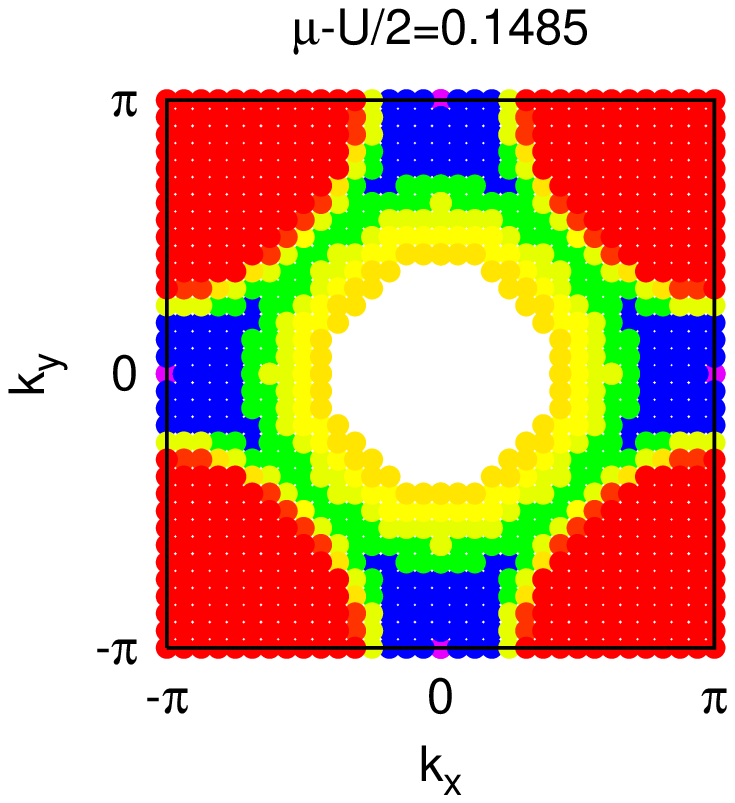}
\includegraphics[scale=0.60, angle=-90]{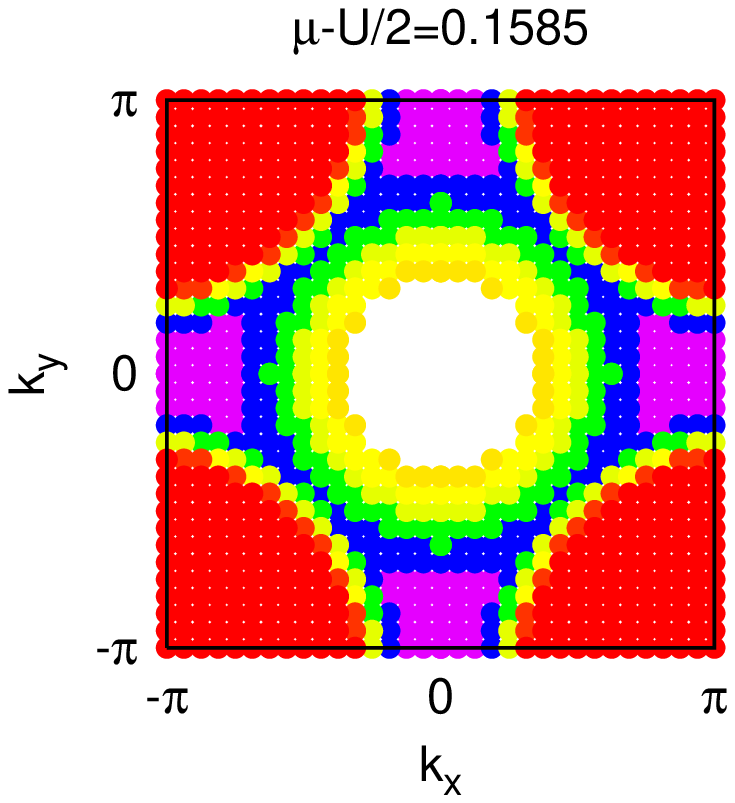}
\end{center}
\includegraphics[scale=0.60, angle=-90]{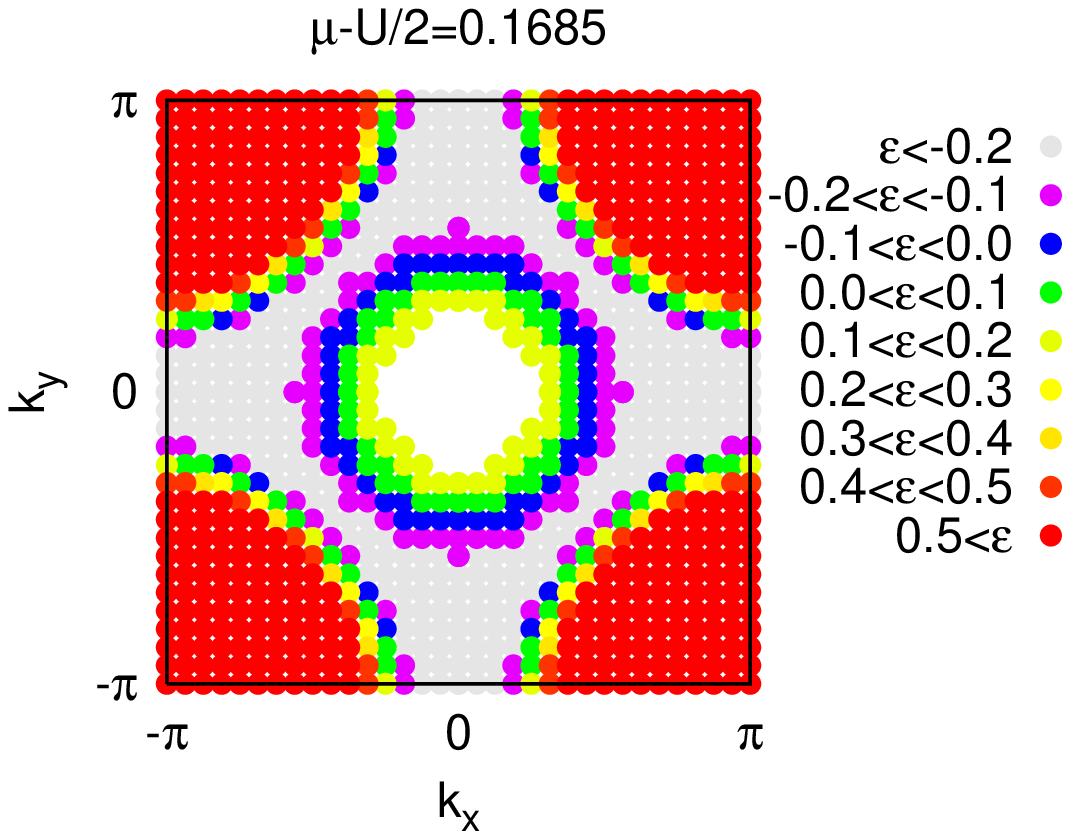}
\caption{%
QP dispersion of the UHB at $U=U_{c}$ for several choices of $\mu$; $\mu-U/2=0.1485$, $0.1585$, and $0.1685$, respectively. The QP pole energy of each k-point is shown with color. The white region corresponds to $k$-points with no sharp QP pole in UHB. Note that blue, purple, and gray regions correspond to $\epsilon_{k}<0$. At $\mu-U/2=0.1585$, the critical point, X' points are at Fermi level, $\epsilon_{k}=0$ and are the topological critical points that connect two hole pockets formed around M-points. 
X'-points are also saddle points of QP dispersion and relatively flat dispersion is found near them. 
}
\label{fig:EDC2Du1050}
\end{figure}

\begin{figure}
\begin{center}
\includegraphics[scale=0.8, angle=-90]{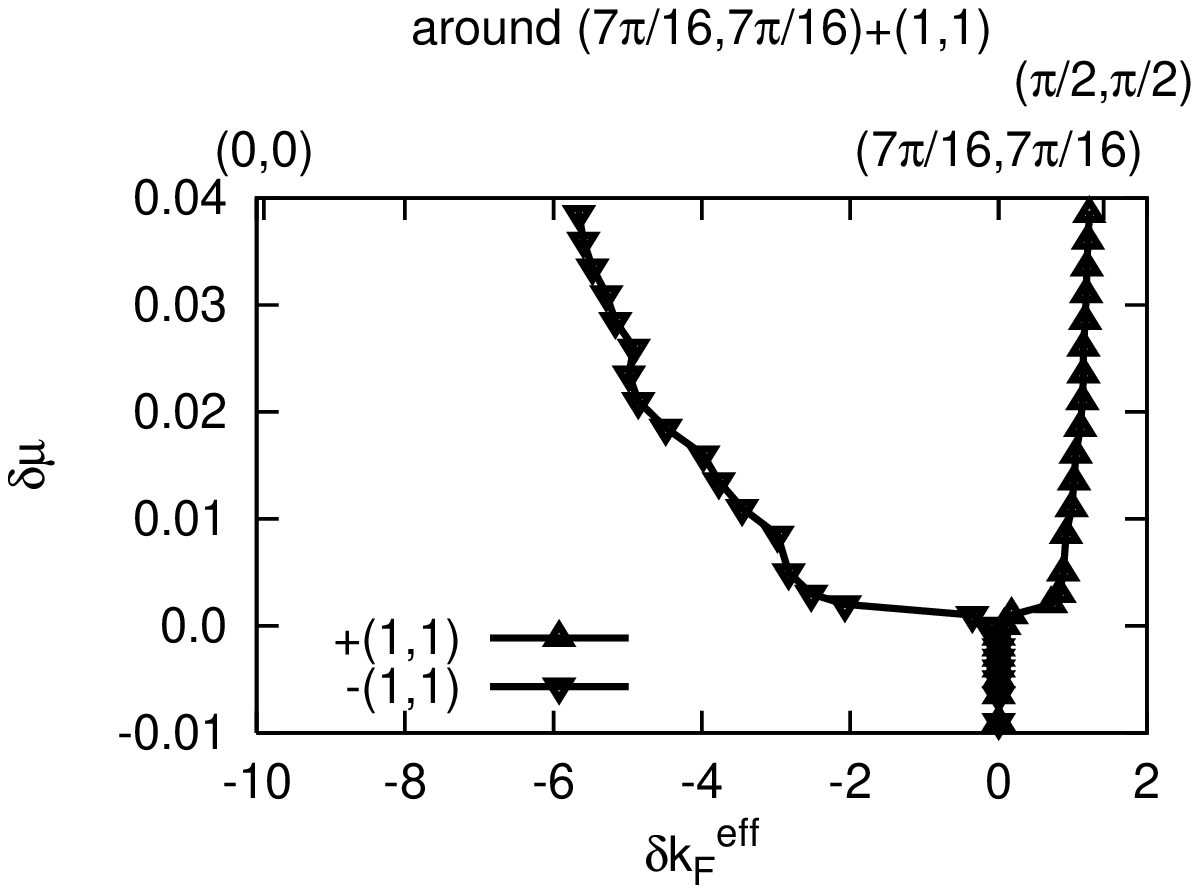}
\includegraphics[scale=0.8, angle=-90]{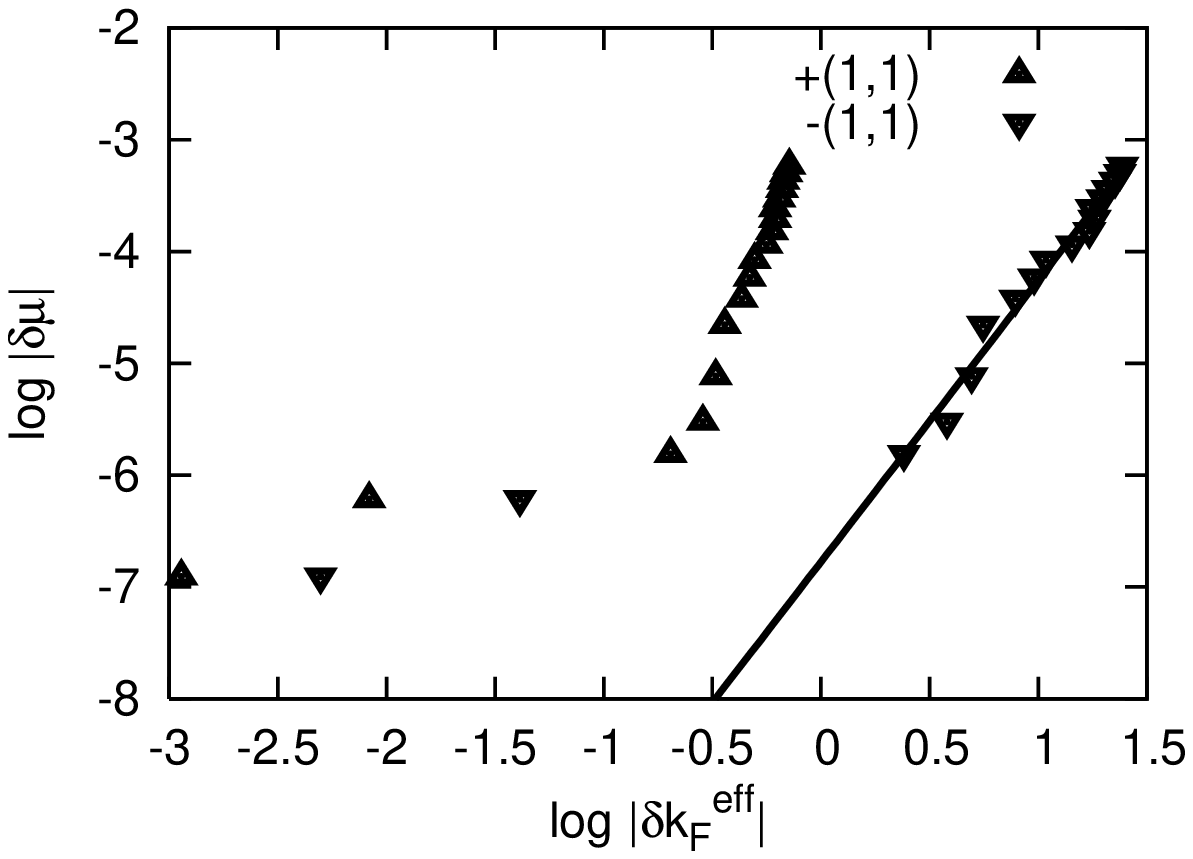}
\end{center}
\caption{
Plot of $\delta k_{F}^{\rm eff}(\delta \mu)$ along $(1,1)$ direction. The origin is set at $k_{F c}=X'=(7\pi/16,7\pi/16)$ and $\mu_{c}-U/2=0.1585$. Top panel shows the plot in the linear scale, while the bottom one shows that in the log-log scale. 
The fitted exponent was $z=2.52$.   
}
\label{fig:krad_u1050}
\end{figure}

In the critical endpoint of the Lifshitz transition at finite temperature, the compressibility diverges as we see above. To get insight into the origin of this divergence of the charge susceptibility $\chi_{c}$ at the Lifshitz critical point, we examine the quasiparticle dispersions. The QP dispersion around the critical point is plotted in Fig.\ref{fig:EDC2Du1050}. It shows a saddle point at $(7\pi/16, 7\pi/16)$, which leads to a flat dispersion near the X-point. We refer to this saddle point and its equivalent points as X$^{\prime}$ points.
The QP dispersion at critical value of $\mu$ around this X$^{\prime}$ point is quadratic in $(1,1)$ direction. 
Thus the divergence of the charge susceptibility is explained within the rigid band picture as the logarithmic divergence at the saddle point;
\begin{equation}
\frac{\partial \langle n \rangle}{\partial \mu}=\frac{1}{N}\int_{\epsilon_{k}=0} dk \frac{1}{|\nabla \epsilon_{k\alpha}|}
\end{equation}
In order to further investigate the flattness of the QP-dispersion around the X$^{\prime}$ points, we plotted the quantity $\delta k_{F}^{\rm eff}(\delta \mu)$ around the X$^{\prime}$ points. 
As we described above, this quantity shows the chemical potential dependence of the evolution of the Fermi surface and expected to be more directly related to the divergence of the charge susceptibility. 
We have derived $k$-exponent $z$ as \( ( \delta k_{F}^{\rm eff} )^{z}=\delta \mu \) as is illustrated in Fig.~\ref{fig:krad_u1050}. Here the obtained $z$ is $z\sim2.5>2$, which also explain the divergence of the compressibility by the saddle point singularity of QP dispersion.
Note that we have fitted the dispersion from $-(1,1)$ direction data only since in that direction, $\delta k_{F}^{\rm eff}(\delta \mu)$ extends more.
We have also discarded data for $\delta k_{F}^{\rm eff}<\sqrt{\mathstrut 2}$, since it depends on the interpolating procedure.~\cite{ref:interpolatingQPEDC} Although we need much larger system size to determine the precise exponent, $z$ larger than 2 obtained within the present accuracy supports the enhancement of flattening of the effective band dispersion beyond the simple van Hove singularity.

\subsection{QP weight around the transition}
\begin{figure}
\includegraphics[scale=0.45,angle=-90]{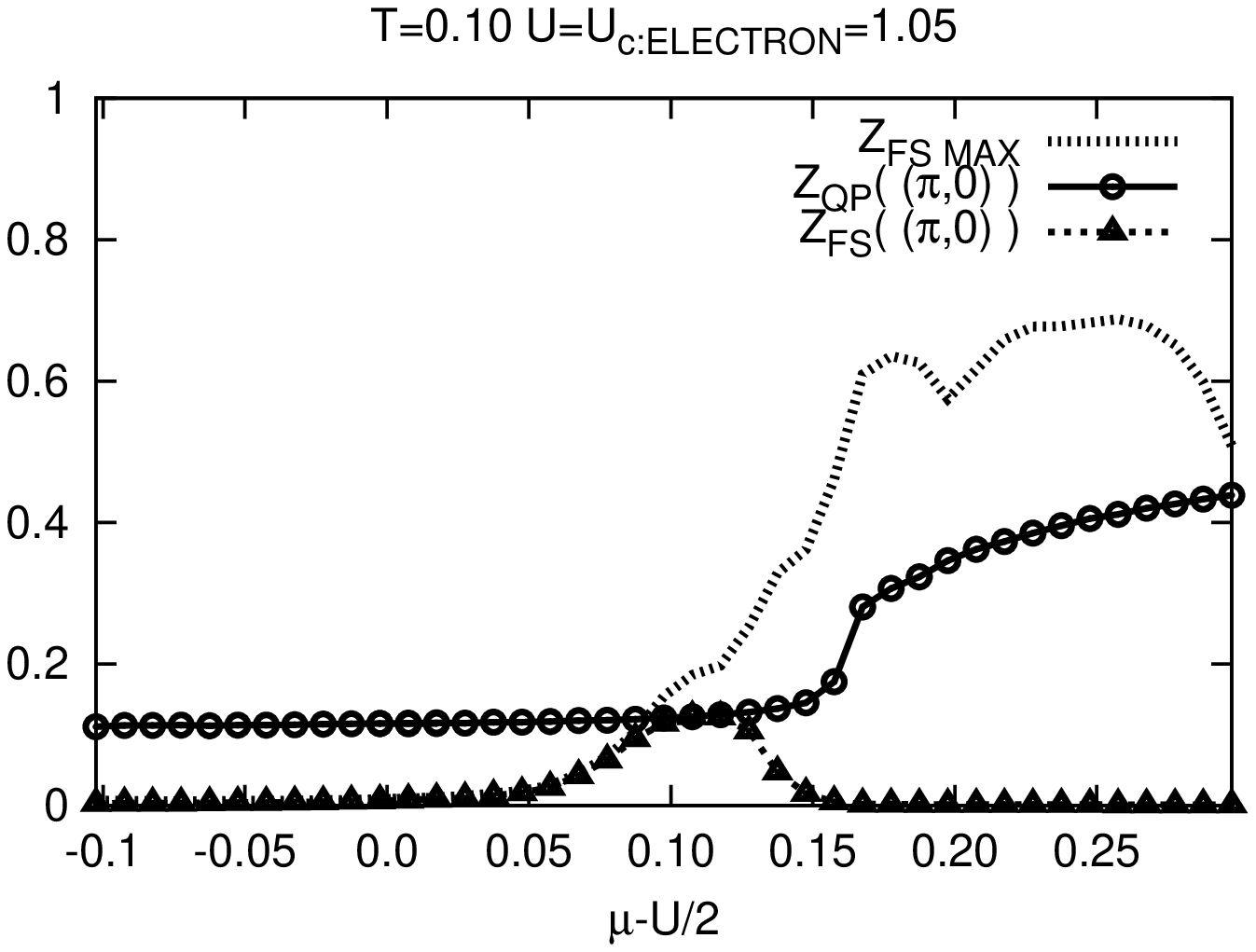}
\includegraphics[scale=0.45,angle=-90]{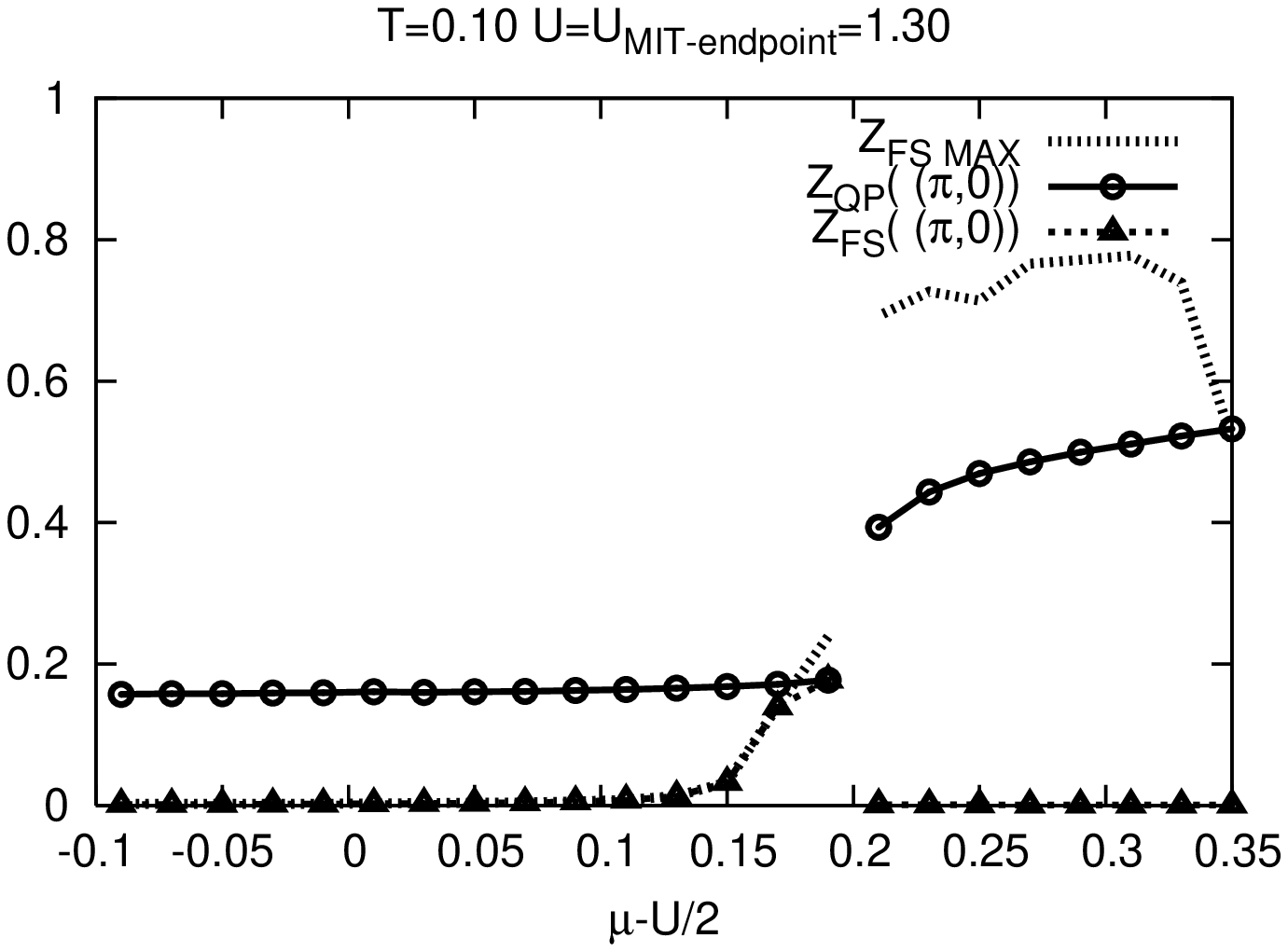}
\includegraphics[scale=0.45,angle=-90]{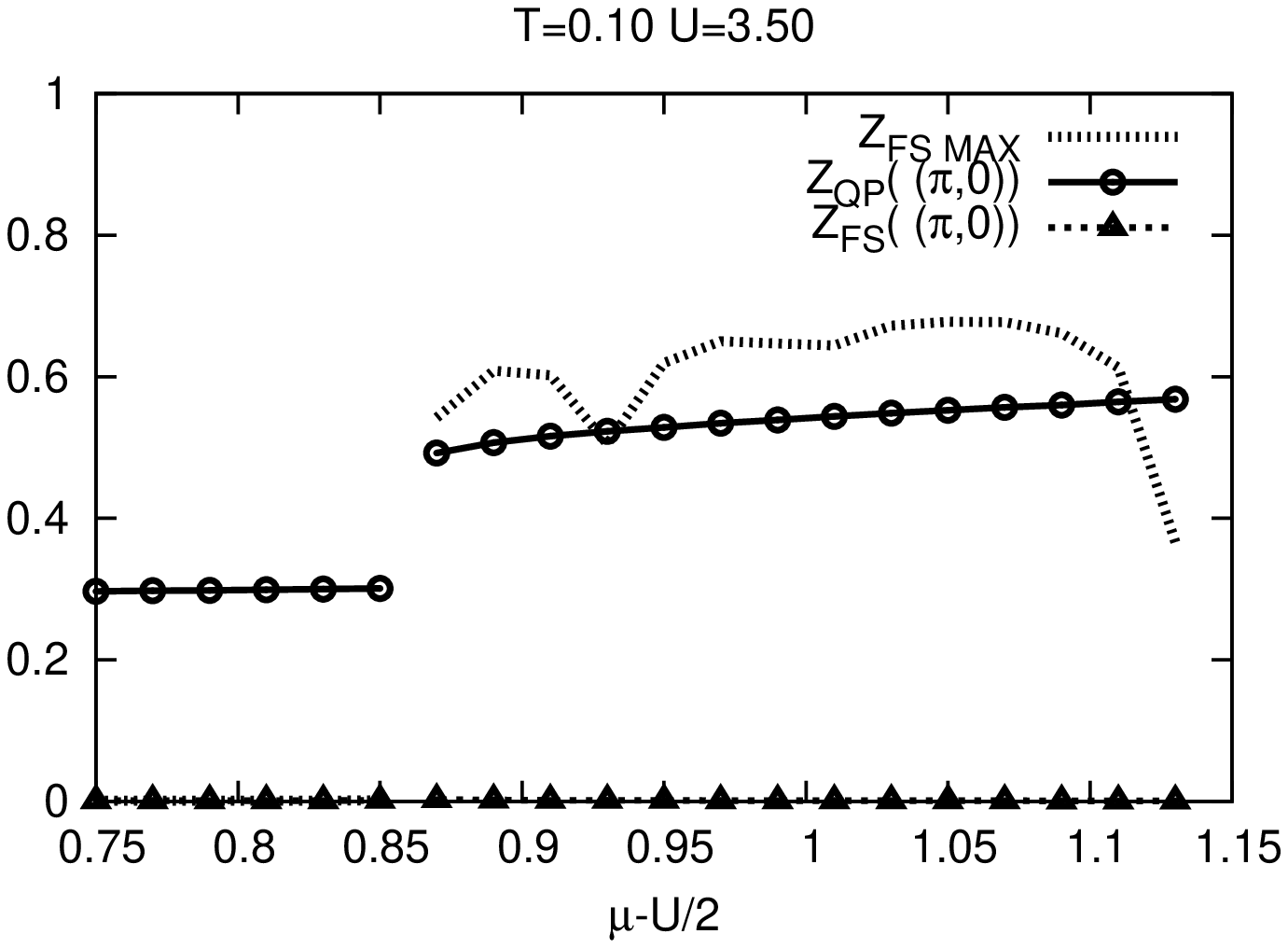}
\caption{%
Plot of QP weights as a function of chemical potential $\mu$. Left-top panel, around the electron-like critical point ( $U=1.05$ ), right-top panel, around the MIT point ($U=1.30$), bottom panel, strong coupling region with discontinuous Lifshitz and metal-insulator transition line ( $U=3.50$ ).
In each panel, largest value of $Z_{\rm FS}$ out of the momentum space, $Z_{\rm FS {\scriptsize \rm MAX} }$, as well as $Z_{QP}$ and $Z_{FS}$ for representative momenta, namely at M point, are plotted against chemical potential $\mu$. 
Note that QP at the M-points is the lowest energy pole of the UHB and play key role for the metal-insulator transition or crossover.
Here, the quantity $Z_{\rm FS {\scriptsize \rm MAX} }$ shows small non-monotonic behavior, though we attribute this to an artifact by finite-size effects coming from discrete momentum points. The overall feature does not suffer from the size effects.
In all the cases, the quasiparticle weight at the M point remains nonzero through the transition.
}
\label{fig:QPWT}
\end{figure}

One of the central issues in this paper is to clarify the nature of the metal-insulator transition and fate of the quasiparticles at the transition. In \S 1, we have introduced two different pictures of MIT. 
In order to clarify the origin and criticality of MIT, we have calculated pole weights of quasiparticles around the critical Lifshitz points as well as around the metal-insulator transition.  
As is shown in Fig. \ref{fig:QPWT}, QP weight at the Fermi level decreases to zero as the system moves from the metallic to the insulating phase. 
Here, we have plotted the change of the QP weights across the transition line at three points with different value of $U$; $U=1.05$, $U=1.30$, and $U=3.50$, which are the electron-like critical point of Lifshitz transition, the end of MIT point, and a point in the strong-coupling region, respectively. The Lifshitz transition is continuous at the first point and of the first order at the latter two. 
Note that in the plot of $U=1.30$, the energy of the lowest energy QP in the UHB becomes zero. QP weight at the Fermi surface remains nonzero even in slightly insulating side of the transition due to finite temperature effects. 

 Clearly, the QP weight at Fermi level vanishes as the system changes to insulating side. 
On the other hand, the QP weights of each QP poles remain finite through the transition. Although the weight jumps at the first-order metal-insulator transition points, the QP weight $Z_{QP}$ remains nonzero in the insulating side even through the first-order metal-insulator transition.  The QP weights $Z_{QP}$ at points other than M point show similar behavior. 

The result indicates that the metal-insulator transition is driven by the shift of Fermi surface out of the QP bands ( LHB or UHB ) rather than vanishing of QP weights irrespective of the order of MIT and irrespective of the involvement of the Lifshitz transition. It supports the {\it quasiparticle rigid-band picture} of MIT rather than the {\it quasiparticle-weight vanishig picture} of MIT. 
QP excitations are still well-defined around the transition although their dispersions change in a non-trivial manner around the transition. Analysis of QP dispersions is crucial for the study of Mott-criticality.   
This result is consistent with quantum Monte Carlo calculation of a single hole in the $t$-$J$ model~\cite{Assaad}.
These results indicate the validity of recently reported QP-based analysis of the Mott criticality.~\cite{ImadaJPSJ1,ImadaJPSJ2,ImadaPRB}
It is remarkable that the higher-order correction of DMFT to include the momentum dependence presented here as CPM seriously modifies the character of the Mott transition from the {\it quasiparticle-weight vanishing} type in the original DMFT to the {\it quasiparticle rigid band} type.

\section{Discussion and Conclusion}
Using CPM, which takes into account momentum dependence of the self-energy beyond the dynamical mean-field theory, we find the Lifshitz transition playing a substantial role near the MIT.
At the Lifshitz transition, the Fermi level crosses the singular point in quasiparticle dispersion. 
This causes a large shift in the QP dispersion and, within a certain parameter region, yields the first-order MIT.
Here, the Lifshitz transition itself causes a large shift in the momentum distribution. However, the self-consistent scheme of CPM relates this shift to the shift in the QP dispersion.
Interestingly, such transition of Fermi surface toplogy is also obtained on the same model, with several different calculation methods. Recent study with the cellular DMFT method reports the change of the Fermi surface topology near half-filling although the relation to MIT has not been studied.~\cite{cDMFTLFST}

However, we also note that all the above results including the present one stays ultimately at the mean-field level of approximations.  By considering the quantum fluctuations more precisely, the first-order transition and the critical temperature might be suppressed.  The compressibility divergence then may occur only at zero temperature in a wide region of the metal-insulator boundary. This is nothing but the marginally quantum critical behavior recently discussed by one of the authors~\cite{ImadaJPSJ1,ImadaJPSJ2,ImadaPRB}. This possibility is supported in the path-integral renormalization group study~\cite{WatanabePIRG}, where the quantum fluctuations are fully taken into account.  
 
Clear Lifshitz transitions are also observed in a recent study with the Hartree-Fock approximation (HFA)~\cite{HFALFST}. In case of HFA, however, it occurs in much more heavily doped (either electron-doped or hole-doped) regime and is rather independent of the metal-insulator transition.
If the Luttinger sum rule~\cite{Luttinger} holds, the critical shape of the Fermi surface determines the critical values of filling. 
When the momentum of the van Hove singularity is far from the folded Broulline-zone boundary for the insulator, the doping concentration required to cause the Lifshitz transition becomes large. This is indeed the case of the Hartree-Fock approximation at large $t^{\prime}/t$. The doping concentration $x_{\rm LFST}$ at the Lifshitz transition is given by the area of the Fermi surface pocket, since the pocket vanishes at the metal-insulator transition.
On the other hand, in CPM, the hole/electron pockets consist of quasiparticles, with a small factor $Z_{QP}$. Therefore the filling at the Lifshitz transition point may become very close to the filling at the MIT (pockets-vanishing) point. They in fact coincide and $x_{\rm LFST}$ becomes zero for larger $U/t$. This may also be interpreted by the renormalization of the Fermi surface shape to the perfectly nested one by the self-energy effect.

When we consider the long-range Coulomb repulsion in addition to the Hubbard model, the first-order transition is suppressed even when the quantum fluctuations are not seriouly considered.  In this situation, the region of the first-order jump in the density (namely, the region of lightly doped Mott insulators at large $U$) becomes a nontrivial phase, where the uniform electron density is unstable but the complete phase separation is suppressed. Quantum fluctuations may again smear out this peculiar phase and lead to a crossover with an unusual metallic behavior.

In conclusion, we have proposed a revised form of CPM to investigate filling-control metal-insulator transition. This formalism of CPM allows incorporating momentum dependence of the self-energy beyond the dynamical mean-field theory with large number of momenta resolved. This makes it possible to study evolution of quasiparticle spectra as well as phase diagram near the Mott transition in detail.  We have studied the filling-control metal-insulator transition in the two-dimensional Hubbard model. Within CPM, we conclude that the quasiparticle survives on the edge of the Mott gap through the metal-insulator transition. The Mott transition is characterized by the Fermi level being separated from the quasiparticle dispersion and entering the gap by keeping nonzero quasiparticle weight.  In this sense, the rigid band picture is ultimately valid on the verge of the transition, where the preformed Mott gap and quasiparticle are both retained through the transition.  At the same time, the actual behavior of the QP dispersion is substantially modified from the original simple rigid band picture.  In a part of the phase diagram, the metal-insulator transition occurs simultaneously with the Lifshitz transition as a combined first-order one. On the other hand, it also shows a region where the two transitions are separated, between which an unusual metallic phase with violation of the Luttinger sum rule appears to be stabilized.     

{\bf Acknowledgements}
The authors thank T. Misawa Y. Yamaji for useful discussions.
This work is supported by a grant-in-aid for scientific research on priority areas from Ministry of Education, Culture, Sports, Science and Technology under the grant numbers 17071003 and 16070612. 


\begin{thebibliography}{999}
\bibitem{ref:KKipt}
In ref. \citen{KajueterKotliar}, they proposed the following two procedures: One is modification of the self-energy $\tilde{\Sigma}_{0}$ obtained by the second order perturbation in the form \( \Sigma=U\langle n_{-\sigma}\rangle +\frac{A\tilde{\Sigma}_{0}}{1-B \tilde{\Sigma}_{0}}. \) 
The other is introduction of the quantity $\tilde{\mu}_{0}$, which shifts the hybridization function $\Delta(\omega)$ and enters in the Weiss field $G_{0}(\omega)$ in the form \( G_{0}(\omega)=\frac{1}{\omega+\tilde{\mu}_{0}-\Delta(\omega)}. \) This quantity $\tilde{\mu}_{0}$ is adjusted by requiring the Luttinger sum rule \(\int_{-i\infty}^{i\infty}d\omega/(2\pi i)G(\omega)\frac{\partial \Sigma}{\partial \omega}.\) See ref.\citen{KajueterKotliar} for detail.
Our idea of adjusting the Weiss field with quantity $\tilde{\mu}$ has similarity to their latter procedure.
\bibitem{ref:interpolatingQPEDC}
Here, in order to obtain a smooth QP energy dispersion curve and its Fermi level section, discrete numerical data of the $32\times32$ k-point mesh are linearly interpolated. Therefore the value of $\delta k^{eff}_{F}$ that comes from less than one k-point mesh spacing is not accurate.
\end{thebibliography}
\end{document}